\documentclass[useAMS,usenatbib,fleqn]{mnras}
\usepackage{graphicx}
\usepackage{amsmath,amssymb}
\usepackage{booktabs}
\usepackage{times}
\usepackage{mathptmx}
\usepackage{url}
\usepackage{epstopdf}
\usepackage{color,soul}

\newcommand{\Mearth}{M_\mathrm{Earth}}
\newcommand{\Msun}{M_\mathrm{Sun}}
\newcommand{\Mmoon}{M_\mathrm{Lunar}}
\newcommand{\Mneptune}{M_\mathrm{Neptune}}
\newcommand{\Rhill}{R_\mathrm{Hill}}
\newcommand{\njs}{\textsc{NJS}}
\newcommand{\ejs}{\textsc{EJS}}
\newcommand{\ejssteep}{\textsc{EJS/Steep}}
\newcommand{\ejsheavy}{\textsc{EJS/Heavy}}
\newcommand{\cjs}{\textsc{CJS}}
\newcommand{\cjssteep}{\textsc{CJS/Steep}}
\newcommand{\cjsheavy}{\textsc{CJS/Heavy}}

\title[Stochasticity \& Predictability in Terrestrial Planet Formation]
      {Stochasticity \& Predictability in Terrestrial Planet Formation}

\author[V. Hoffmann et al.]
       {\parbox[t]{\textwidth}
        {Volker Hoffmann,$^{1}$\thanks{E-mail: volker@physik.uzh.ch}
         Simon L. Grimm,$^{1,2}$
         Ben Moore$^{1}$
         and Joachim Stadel$^{1}$}
         \vspace*{6pt}\\
        $^{1}$Institute for Computational Science,
            University of Z\"urich,
            CH-8057 Z\"urich, Switzerland\\
        $^{2}$Center for Space and Habitability,
            University of Bern,
            CH-3012 Bern, Switzerland
        \vspace*{6pt}}

\begin{document}

\date{Accepted 03 November 2016.
      Received 03 November 2016; in original form 04 August 2015}

\pagerange{\pageref{firstpage}--\pageref{lastpage}}

\pubyear{2016}

\maketitle

\label{firstpage}

\begin{abstract}
Terrestrial planets are thought to be the result of a vast number of gravitational interactions and collisions between smaller bodies. We use numerical simulations to show that practically identical initial conditions result in a wide array of final planetary configurations. This is a result of the chaotic evolution of trajectories which are highly sensitive to minuscule displacements. We determine that differences between systems evolved from virtually identical initial conditions can be larger than the differences between systems evolved from very different initial conditions. This implies that individual simulations lack predictive power. For example, there is not a reproducible mapping between the initial and final surface density profiles. However, some key global properties can still be extracted if the statistical spread across many simulations is considered. Based on these spreads, we explore the collisional growth and orbital properties of terrestrial planets which assemble from different initial conditions (we vary the initial planetesimal distribution, planetesimal masses, and giant planet orbits). Confirming past work, we find that the resulting planetary systems are sculpted by sweeping secular resonances. Configurations with giant planets on eccentric orbits produce fewer and more massive terrestrial planets on tighter orbits than those with giants on circular orbits. This is further enhanced if the initial mass distribution is biased to the inner regions. In all cases, the outer edge of the system is set by the final location of the $\nu_6$ resonance and we find that the mass distribution peaks at the $\nu_5$ resonance. Using existing observations, we find that extrasolar systems follow similar trends. Although differences between our numerical modelling and exoplanetary systems remain, we suggest that CoRoT-7, HD 20003, and HD 20781 may host undetected giant planets.
\end{abstract}



\section{Introduction}

Numerical simulations suggest that our solar system is inherently chaotic and that any small changes to the planets' initial positions diverge exponentially with e-folding times on the order of $5$ to $20$ $\mathrm{Myr}$ \citep{1988Sci...241..433S,1989Natur.338..237L,1991AJ....101.2287Q,1992Sci...257...56S,1994A&A...287L...9L}. However, the difference between chaotic and near-integrable orbits remains within the bounds of measurement uncertainties of the outer planets \citep{2008MNRAS.386..295H}. Although the outer solar system planets appear to be remarkably stable against developing crossing orbits on $\mathrm{Gyr}$ timescales \citep{2002MNRAS.336..483I}, this may not be the case for the inner solar system planets \citep{2008Icar..196....1L,2009Natur.459..817L}.

On such timescales, chaos is mediated by overlapping resonances \citep{1998AJ....116.3029N,1997AJ....114.1246M,1995Icar..114...33M,1993Icar..102..316M,1992Icar...95..148L,1990Icar...88..266L,1984prin.conf..562F,1980AJ.....85.1122W,1979PhR....52..263C} to which minor bodies are particularly sensitive. Analytical and numerical work suggests that overlapping resonances account for the observed distribution of bodies in the asteroid belts \citep{1998Icar..135..458M,1997Sci...277..197G,1985Icar...63..272W,1983Icar...56...51W,1982AJ.....87..577W}, in the inner \citep{1999Natur.399...41E,1995MNRAS.277..497M} and outer solar system \citep{2005Natur.435..462M,1999Icar..140..341G,1999Icar..140..353G,1993AJ....105.1987H,1995ceme.symp..116H,1993ARA&A..31..265D,1990AJ....100.1680G,1989Icar...79..223F,1973AJ.....78..329E,1973Icar...20..422L}, as well as within the Kuiper belt \citep{1997Sci...276.1670D,1997Icar..127...13L,1993AJ....105.1987H,1993ApJ...406L..35L,1990Natur.345...49T,1989AJ.....98.1477T}.\footnote{An extensive overview on resonances is given in \cite{2001ARA&A..39..581L}.}



During the epoch of terrestrial planet formation, the solar system environment was rather different than today. Set against a backdrop of migrating giant planets \citep{2011Natur.475..206W,2011AJ....142..152L,2007Icar..191..158M,2007AJ....134.1790M,2005Natur.435..459T} and being embedded in a dissipating gaseous disk \citep{2014ApJ...793L..34P,2009AIPC.1158....3M,2001ApJ...553L.153H}, planetesimals are thought to grow collisionally and hierarchically into terrestrial planets \citep{2000Icar..143...15K,1998Icar..131..171K,1997Icar..128..429W,1996Icar..123..180K,1992Icar...98...28I,1992Icar...96..107I,1992Icar..100..440G,1990Icar...87...40G,1989Icar...77..330W,1978Icar...35....1G,1969Icar...10..109S}. Planetesimals undergo perturbational encounters with each other (to within a few Hill radii) about once per orbit, whereas perturbations from resonant configurations with giant planets require hundreds of orbital periods to cause noticeable effects. Planetesimal disk dynamics resemble those of stellar systems, in which small orbital perturbations grow exponentially fast \citep{2002JSP...109.1017H,2000chun.proc..229V,1993ApJ...415..715G,1991ApJ...374..255K,1964ApJ...140..250M}. This is the essence of stochasticity in planetesimal disks.

Numerical simulations probe this regime by tracking the collisional evolution of planetesimals. Such simulations can address the formation and composition of the terrestrial planets \citep{1998Icar..136..304C,2001Icar..152..205C,2004Icar..168....1R,2005ApJ...632..670R,2006ApJ...642.1131K,2006Icar..184...39O,2006Icar..183..265R,2009Icar..203..644R,2010Icar..207..517M} and the extrasolar systems \citep{2014ApJ...787..172O,2014ApJ...794...11I,2006ApJ...644.1223R,2005Icar..177..256R}.\footnote{The literature on terrestrial planet formation can be overwhelming. Excellent reviews include \cite{2014prpl.conf..595R} and \cite{2012AREPS..40..251M}.} 

In these simulations, much of the dynamics of asteroids, planetesimals, and embryos in the inner solar system depends on the presence of giant planets, with interactions mediated most strongly by way of mean-motion and secular resonances. The latter depend on the gravitational potential of the gas disk, so that their location sweeps inwards as the gas dissipates, which provides a credible mechanism to dynamically excite bodies in large parts of the inner solar system. Attempting to reproduce the distribution of eccentricities and inclinations in the asteroid belt, \cite{1980Icar...41...76H}, \cite{1981Icar...47..234W}, \cite{1997Icar..129..134L}, \cite{2000AJ....119.1480N}, and \cite{2001EP&S...53.1085N} demonstrated the mechanism of resonance sweeping on massless test particles. In the context of terrestrial planet formation, \cite{1998Icar..136..304C} and \cite{2001Icar..152..205C} numerically demonstrated the effect of secular interactions of embryos with Jupiter and Saturn (as well as among the embryos themselves), although their experiments did not consider a gaseous disk. Later, \cite{2003AJ....125.2692L} investigated the effect of various giant planet configurations on planetary embryos and described ``secular conduction'' which allows embryos at secular resonances to excite embryos in other regions. Adding gas, \cite{2004Icar..167..231K} numerically found that including of a decaying gas disk and giant planets leads to significantly shorter crossing times. Finally, \cite{2005ApJ...635..578N} and \cite{2008ApJ...676..728T} demonstrated that inward secular resonances sweeps along embryos to the inner systems as gas dissipates. Extending preceding work, \cite{2010Icar..207..517M} demonstrated that the secular resonances also sweep planetesimals inward. In all cases, authors stress that the tendency of sweeping resonances to dynamically excite embryos and planetesimals is balanced by the dampening influence of hydrodynamic drag and dynamical friction (sometimes called gravitational drag).

Irrespective of the detailed setup of simulations, their initial conditions are usually generated by drawing realisations from some underlying solid mass distribution, possibly subject to stability constraints if planetary embryos are implanted directly \citep{1996Icar..119..261C,1999Icar..139..328Y}. As the system is inherently chaotic, we expect that different initial conditions drawn from the same underlying distribution will lead to different final systems, much like in simulations of stellar dynamics \citep{2012MNRAS.424..272P,2010MNRAS.407.1098A}. To date, few contributors have evolved multiple realisations of the same distribution as limited computational resources are typically focussed on parameter studies.\footnote{Instead of parameters sweeps, \cite{2000Icar..143...45R} tackled numerical issues and pushed the number of massive particles to $10^6$.} Those that did report distinctly different outcomes for different realisations of the same underlying distribution \citep{2014ApJ...782...31I,2011Natur.475..206W,2006ApJ...642.1131K,2009Icar..203..644R}. Therefore, just how reliable are simulations of the collisional growth of terrestrial planets -- do they have predictive power?

This question is at the heart of our paper and we tackle it in a twofold manner. First, we evolve identical realisations of a planetesimal disk multiple times. Due to differences in round-off errors, we will see that simulations terminate with different planetary configurations and we assess the statistical spread of several diagnostics. Second, we evolve the planetesimal disk in the absence and presence of Jupiter and Saturn (in two configurations). We also vary the initial mass and planetesimal distribution. Again, we compare diagnostics across runs and check whether trends in the diagnostics are visible or buried in the statistical spread.

The paper is organised as follows: In Section \ref{sec:methods}, we outline numerical methods, initial conditions, the analytic gas model used, and describe our simulations. In Section \ref{sec:results}, we present results by covering some illustrative examples, describing how the planetesimal disks evolve, what kind of final systems they lead to, describing the driving dynamics, determining whether a simple mapping between initial and final surface density profile persists, and addressing variations in typical diagnostics used in terrestrial planet formation. In Section \ref{sec:discussion}, we extend this by discussing caveats in our dynamical modelling and attempting to match our trends to observations. We conclude the paper in Section \ref{sec:conclusions}.


\section{N-Body Methods, Initial Conditions}
\label{sec:methods}

In this work, we use the GPU code \textsc{Genga} to follow the collisional evolution of planetesimal disks. We now describe the code, skirt the issue of program execution order, the initial conditions of the planetesimal disk, and the gas disk model.

\subsection{\textsc{Genga}}

\textsc{Genga} \citep{2014ApJ...796...23G} is a hybrid symplectic integrator similar to \textsc{Mercury} \citep{1999MNRAS.304..793C}, but running in parallel on Graphics Processing Units (GPUs). The integration scheme treats gravitational interactions between bodies as perturbations of their Keplerian orbits. \textsc{Genga} uses democratic coordinates (heliocentric positions, barycentric velocities) \citep{1998AJ....116.2067D}. This allows the code to separate close encounter pairs from the rest of the system, and integrate them separately with a direct N-Body integrator up to machine precision. Outside of close encounters, the bodies are integrated with a symplectic integrator. The hybrid symplectic integrator has excellent energy conservation over a large number of orbits. Accelerations between bodies are computed directly. This requires $\mathcal{O}(N^2)$ operations, which is more efficiently calculated on a GPU and is more accurate than a tree-based method. \textsc{Genga} supports an analytical gas disk model inherited from a patched version of \textsc{Pkdgrav} \citep{2001PhDT........21S,2010Icar..207..517M} which we describe in Section \ref{ssec:ic}. The code is available online.\footnote{\url{https://bitbucket.org/sigrimm/genga}}

\subsection{Forcing the Order of Program Execution}
\label{ssec:serial_grouping}

To exactly reproduce numerical results, the order of execution of steps within the program must be fixed.\footnote{In computer arithmetic, $a+b+c \neq a+c+b$ because storage space for each number is finite, such that round-off errors will  differ.} While this is easily controlled in single threaded applications, multi-threaded programs require additional logic. To ensure reproducibility of numerical experiments and probe the underlying mechanics of orbital divergence, we have implemented such logic in \textsc{Genga}.

The most likely source of variations in the order of operations is the parallel sum operation. In \textsc{Genga}, this is implemented as a parallel reduction formula within  one thread block and always operates in the same order. As such, all summation operations are excluded as the source of round-off error variations. The only remaining possible source is in the creation of the close encounter list. If a close encounter pair is found, a counting variable is increased through an \texttt{\_atomicAdd()} operation. The order of this operation is not well defined across threads. A different order of the close encounter pair list leads to a different order of bodies in the direct N-Body integrator. For multiple close encounter groups, this can lead to a different result. We prevent this behaviour through an additional sorting step, which reorders the close encounter list, but induces a performance penalty. The behaviour is controlled at compilation through the \texttt{SERIAL\_GROUPING} flag.

In this work, all simulations run with this flag disabled because we rely on variations in round-off errors to induce orbital divergence. Tests show that individual runs of Section \ref{sec:results} can be reproduced exactly if we enable \texttt{SERIAL\_GROUPING}. We are presently preparing a companion paper which exploits this flag to determine the rate of orbital divergence.

\subsection{Initial Conditions, Gas Disk}
\label{ssec:ic}

\renewcommand{\arraystretch}{1.3}
%
%
\begin{table*}
\caption{Orbital elements of Jupiter and Saturn as well as initial planetesimal disk conditions for our simulation sets. For the giant planets, the three angular arguments are initialised to zero. The \cjs set corresponds to the initial conditions of the Nice model \citep{2005Natur.435..459T}. The \ejs{} set corresponds to the present-day solar system. In all cases, we start with $2000$ planetesimals.} 
\label{table:simulation_setup}
\begin{center}
\begin{tabular}{c c c c c c c c c c c c}
\toprule

  Set
& $a_\mathrm{J}$ (AU) \textsuperscript{a}
& $e_\mathrm{J}$ \textsuperscript{b}
& $i_\mathrm{J}$ (Deg) \textsuperscript{c}
& $a_\mathrm{S}$ (AU) \textsuperscript{a}
& $e_\mathrm{S}$ \textsuperscript{b}
& $i_\mathrm{S}$ (Deg) \textsuperscript{c}
& $\Sigma_\mathrm{Disk}$ \textsuperscript{d}
& $\Sigma_{\mathrm{Disk},0}$ (g/cm$^2$)  \textsuperscript{e}
& $M_\mathrm{Disk}$ ($\Mearth$) \textsuperscript{f}
& $N_\mathrm{Runs}$ \textsuperscript{g} \\

\midrule

  \njs
& --
& --
& --
& --
& --
& --
& $\varpropto r^{-1}$
& $6.1$
& $5$
& $12$ \\

  \ejs
& $5.2$
& $0.048$
& $1.30$
& $9.55$
& $0.056$
& $2.49$
& 
& 
& 
& \\

  \cjs
& $5.45$
& $0.0$
& $0.0$
& $8.18$
& $0.0$
& $0.5$
& 
& 
& 
& \\

\midrule

  \ejssteep
& $5.2$
& $0.048$
& $1.30$
& $9.55$
& $0.056$
& $2.49$
& $\varpropto r^{-1.5}$
& $8.2$
& $5$
& \\

  \ejsheavy
& 
& 
& 
& 
& 
& 
& $\varpropto r^{-1}$
& $12.4$
& $10$
& \\

\midrule

  \cjssteep
& $5.45$
& $0.0$
& $0.0$
& $8.18$
& $0.0$
& $0.5$
& $\varpropto r^{-1.5}$
& $8.2$
& $5$
& \\

  \cjsheavy
& 
& 
& 
& 
& 
& 
& $\varpropto r^{-1}$
& $12.4$
& $10$
& \\

\bottomrule
\end{tabular}
\end{center}
\raggedright
\textsuperscript{a} Semi-Major Axis.
\textsuperscript{b} Eccentricity.
\textsuperscript{c} Inclination.
\textsuperscript{d} Surface Density Profile.
\textsuperscript{e} Surface Density at $1 \mathrm{AU}$.
\textsuperscript{f} Disk Mass.
\textsuperscript{g} Number of Independent Runs.\\
\end{table*}
\renewcommand{\arraystretch}{1.0}

Initial conditions (ICs) are generated the same way as in \cite{2010Icar..207..517M}, where a number of samples (planetsimals of equal mass) are drawn from an underlying distribution of Keplerian elements. This generates a realisation with a particular surface density profile and total mass. We generate realisations by drawing $2000$ samples such that the radial surface density profile is

\begin{equation}
  \Sigma_\mathrm{Disk}(r) = 
  \begin{cases}
    \Sigma_{\mathrm{Disk},0} \left( \frac{r}{1 \, \mathrm{AU}}\right)^{-p} \qquad 0.5 \; \mathrm{AU} < r < 4 \; \mathrm{AU}, \\
    0 \qquad \qquad \qquad \qquad \quad \mathrm{otherwise},
  \end{cases}
\end{equation}

for total disk mass of

\begin{equation}
  M_\mathrm{Disk} = 2 \pi \int_{0.5}^4 \Sigma(r) \, r \; \mathrm{d}r.
\end{equation}

In this paper, we adopt $p = 1$ and $1.5$ as well as $M_\mathrm{Disk} = 5 \; \Mearth$ and $10 \; \Mearth$ (see Section \ref{ssec:setup} and Table \ref{table:simulation_setup}). All planetesimals have mass $M \sim 0.04 \; \Mmoon$ ($\sim 0.08 \; \Mmoon$ for $10 \; \Mearth$ disks) and are on nearly circular ($e < 0.02$), low inclination ($i<0.75^\circ$) orbits. The planetesimals are embedded in a gas disk described by an analytical model. The gas surface density follows a power law and decays exponentially in time, i.e.

\begin{equation}
  \label{eq:gasdisk}
  \Sigma_\mathrm{Gas}(r,t) = \Sigma_{\mathrm{Gas},0} \left( \frac{r}{1 \, \mathrm{AU}} \right)^{-1} \exp \left( - \frac{t}{\tau} \right),
\end{equation}

where $\tau$ is the decay time of the gas disk. For all simulations, $\tau = 1 \; \mathrm{Myr}$, $\Sigma_{\mathrm{Gas},0} = 2000 \; \mathrm{g}/\mathrm{cm}^2$. After $\sim 4.6 \; \mathrm{Myr}$, only $1\%$ of the gas remains.

Particles exchange angular momentum with the gas disk in three ways: (i) hydrodynamic drag due to differences in velocity, (ii) torques arising from spiral density waves launched by massive particles, and (iii) gravitational interactions between particles and the massive disk. Note that we artificially enhance hydrodynamic drag for particles with masses $< 0.01 \; \Mearth$ to correct for the large initial planetesimal mass. For more details on the interaction between gas and particles, we refer to Section 2.2 in \cite{2010Icar..207..517M}, which we have implemented verbatim. We stress that these interactions are modelled analytically. Our simulations would benefit from a full hydrodynamic model,\footnote{In particular, spiral density waves launched from multiple massive bodies will locally modify the hydrodynamic drag, mutually interact, and affect the gravitational potential of the disk. Neither of these effects are captured in an analytic model.} which may affect some of the results in this work.

\subsection{Simulation Setup, Post-Processing}
\label{ssec:setup}

We generate planetesimal disks by drawing a single realisation for three different distributions: (i) a reference disk, (ii) a disk with a steeper surface density profile, and (iii) a more massive disk. Note that even for steeper planetesimal disk profiles, we retain a gaseous disk with suface density profile $\sim r^{-1}$, which is consistent with models of dust coagulation \citep{2012A&A...539A.148B}. For each realisation, we generate two sets of simulations with giant planets -- one with Jupiter and Saturn on circular orbits (\cjs) and one with Jupiter and Saturn on eccentric orbits (\ejs). The \ejs{} set corresponds to the present-day solar system and the \cjs{} set to the initial conditions of the original Nice model \citep{2005Natur.435..459T}. The Jovian planets are inserted at the start of the simulations. We also generate a set based on the reference disk without giant planets (\njs). Table \ref{table:simulation_setup} summarises our initial conditions.

Each set is evolved $12$ times for a total of $84$ runs covering 9 billion steps ($t \sim 147.84 \; \mathrm{Myr}$, $\Delta t = 6 \; \mathrm{days}$). Computing time per run is about a month on a NVIDIA GeForce GTX 590. We treat particle collisions as inelastic mergers. Particles are removed from the simulations if their heliocentric distance falls below $0.2 \; \mathrm{AU}$ or exceeds $20 \; \mathrm{AU}$. Given our six day timestep, this makes for at least $5.4$ steps per orbit. The relative energy error remains $\Delta E/E_0 < 4.3 \times 10^{-4}$ at all times. All the outputs are available online.\footnote{\url{https://cheleb.net/astro/sp15}}

In post-processing, we load the outputs from all twelve runs per set, calculate various diagnostics for each run, and plot/tabulate their median, as well as $10/90$ and $25/75$ percentile spreads across the twelve runs. In the text, we quote only median values and $10/90$ percentile ranges.

For simulations with giant planets we compute the location of the $\nu_5$, $\nu_6$, $\nu_{15}$, and $\nu_{16}$ secular resonances to overlay during the analysis. The calculation is implemented as in the appendix of \cite{2000AJ....119.1480N} with the following caveats: (i) the locations are exact only for massless test particles with on low eccentricity/inclination orbits, (ii) they are computed to first order and therefore only depend on the semi-major axis $a$ (as well as the masses and semi-major axes of the giants), (iii) the gas disk is modelled as in Eqn. \eqref{eq:gasdisk}, ignoring modifications of the potential by spiral density waves (whose effect is modelled analytically). We thus expect the location of the resonances in the simulation to slightly differ from the values derived in post-processing. Also note that the $\nu_{15}$ resonance does not appear in the region of interest, and that -- as the gas dissipates -- the $\nu_{16}$ resonance appears in two locations. See also Figure 4 in \citet{2000AJ....119.1480N}.

For consistency, we apply the same colour-coding of simulation sets to all figures in this paper. Blue colours indicate \njs{} runs, red colours \ejs{} runs, and green colours \cjs{} runs. Depending on the context, the shade encodes different information. If only one run is shown (or multiple runs are aggregated in an otherwise discernable fashion), the shading correlates with the mass. If multiple runs are aggregated to describe statistics of simulation sets, the shading indicates the statistical spread, i.e. median values as well as relevant percentile offsets.


\section{Results}
\label{sec:results}

We now present results from the runs described above. We begin by illustrating how two identical initial conditions diverge rapidly and lead to different systems as well as the principal mechanism that drives our planetesimal disks. Afterwards, we address the assembly of embyros from planetesimals, final system architectures, radial mass and mass-frequency distributions, the link between planetesimal distributions and system architecture, the long-term stability of our systems, how resonances sculpt the system, and the spread in frequently used diagnostics.


\subsection{Some Illustrative Examples}

\subsubsection{Divergence of Orbits \& Runs}

\begin{figure}
\centering
\includegraphics[scale=0.86]{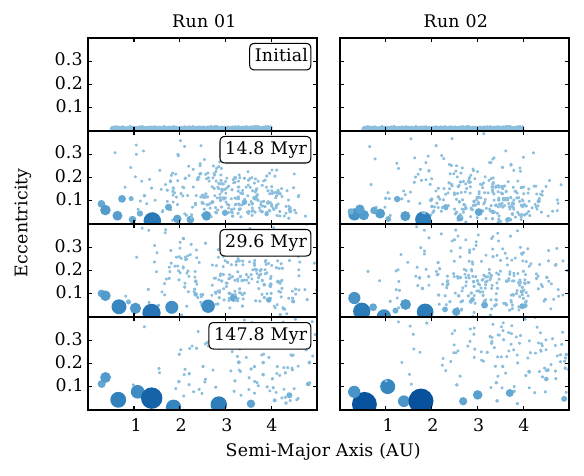}
\caption{Semi-major axis and eccentricity for two identical initial conditions (columns) at four time slices (rows). Each marker represents a single body. Larger markers indicate more massive bodies.}
\label{fig:fig01_aesnaps_small}
\end{figure}

Figure \ref{fig:fig01_aesnaps_small} shows a time sequence of the semi-major axis and eccentricity for two of the twelve simulations launched without Jupiter and Saturn. By $14.98 \; \mathrm{Myr}$, the simulations have clearly evolved different groups of planetary embryos as well as populations of remaining planetesimals. After $147.84 \; \mathrm{Myr}$, the simulations terminate with distinctly different terrestrial planets.

Closer analysis reveals that initially identical orbits diverge exponentially fast with e-folding times on the order of a few to a few tens of years. Although the particle position and velocity vectors are initially identical, variations in round-off errors across simulation runs induce position variations at the level of floating point accuracy (one part in $10^{15}$, or about a millimetre for planets at $1 \; \mathrm{AU}$) which grow exponentially due to the chaotic nature of the gravitational N-Body problem -- see, e.g., \cite{1993ApJ...415..715G}. On timescales of a few hundred years, two initially identical simulations diverge fully and the systems will undergo different collisional histories. The rate of divergence of initially nearby planetesimals depends on the mass resolution, i.e. the number of particles initialised to sample a given total disk mass. In general, increasing the number of particles accelerates divergence. This has potential implications for the resolution requirements. We cover this in a companion paper.

\subsubsection{Sweeping Secular \& Mean Motion Resonances}
\label{sssec:sweeping_resonances}

\begin{figure*}
\centering
\includegraphics[scale=0.96]{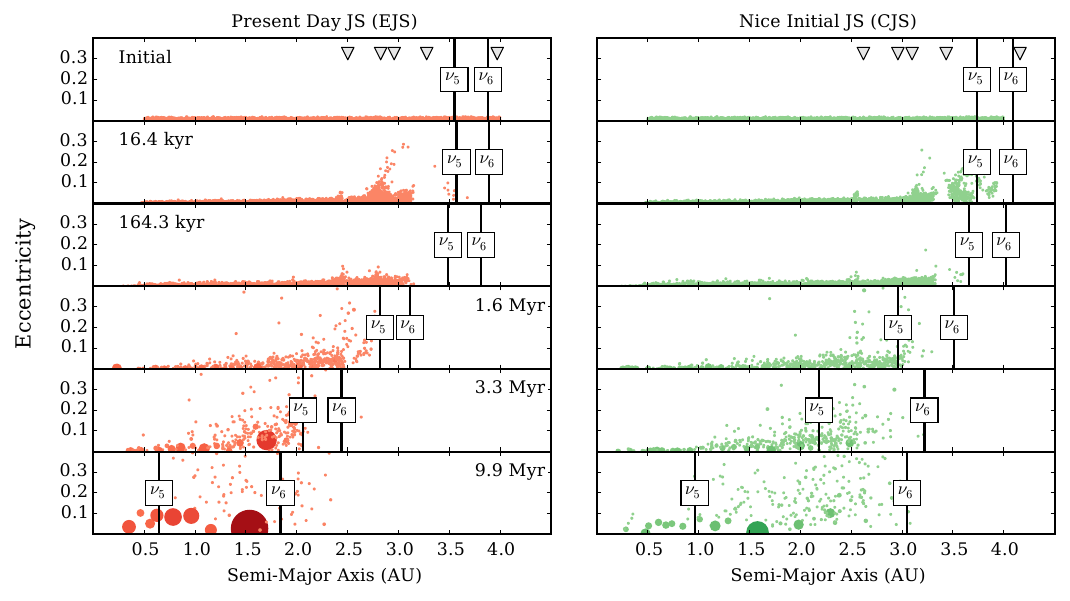}
\caption{Semi-major axis and eccentricity for a single run with \ejs{} (left) and \cjs{} (right) conditions at six time slices (rows). Markers represent single bodies with size is proportional to the mass. Triangles indicate the location of (from left to right) 3:1, 5:2, 7:3, 2:1, and 3:2 mean motion resonances with Jupiter. Vertical bars indicate the location of the secular resonances $\nu_5$ and $\nu_6$, which sweep inwards as the gas disk dissipates. We do not show the $\nu_{16}$ resonance which settles $\sim 0.1 \; \mathrm{AU}$ beyond the $\nu_6$ resonance at $\sim 9.9 \; \mathrm{Myr}$. Animations are available at \protect\url{https://cheleb.net/astro/chaos15/media/}.}
\label{fig:fig02_ae_snaps_res}
\end{figure*}

Most of our runs host giant planets which primarily interact with the planetesimal population by way of mean motion and secular resonances. While mean motion resonances remain fixed in space (up to orbital variations in the giant planets), secular resonances can sweep through the regions populated by planetesimals as the protoplanetary gas disk dissipates. Doing so, they may shepherd planetesimals along to sculpt the final architecture of the terrestrial planets \citep{2005ApJ...635..578N,2008ApJ...676..728T}. Postponing discussion on the influence of sweeping secular resonances, we now recapitulate their origin and illustrate how they influence planetesimal dynamics.

Due to their mutual gravitational interaction, the Jupiter-Saturn system has four eigenfrequencies labelled $f_1$, $f_2$, $g_1$, and $g_2$. Linear combinations of these drive secular variations in eccentricity and inclination \citep{1999ssd..book.....M}. They depend on the planetary masses, orbital configuration, and gravitational potential of the gas disk in which they are embedded \citep{1961mcm..book.....B,1980Icar...41...76H,1981Icar...47..234W,2000AJ....119.1480N}. Given interactions with Jupiter and Saturn as well as the gravitational potential of the gas disk, planetesimals are subjected to perturbations with frequencies $f$ and $g$. At particular locations in the disk, these frequencies match, driving resonances — labelled $\nu_5$ ($g = g_1$), $\nu_6$ ($g = g_2$), $\nu_{15}$ ($f = f_1$), and $\nu_{16}$ ($f = f_2$). They pump eccentricities ($\nu_5$, $\nu_6$) or inclinations ($\nu_{15}$, $\nu_{16}$).

Figure \ref{fig:fig02_ae_snaps_res} shows the location of the secular resonances at different timeslices. As time moves on, the gas dissipates and the secular resonances sweep inwards, pushing planetesimals in front of them. For the \ejs{} configuration, $\nu_5$ and $\nu_6$ remain closer together than for the \cjs{} case. In the \ejs{} case, $\nu_5$ also settles at smaller semi-major axes $a$ once the gas disappears. The net effect is that more material is delivered to smaller $a$ (where growth timescale are faster) in \ejs{} configurations. As we will confirm below, this leads to faster assembly and growth of planetary embryos.

The mean motion resonances occur at semi-major axes where planetesimals and giant planets periodically line up at fixed orbital phases. We indicate five of the lowest order mean motion resonances by triangles in Figure \ref{fig:fig02_ae_snaps_res}. As the simulations evolve, their location remains approximately fixed. At early times, they are efficient at exciting planetesimals, although eccentricities are rapidly dampened by hydrodynamic drag. As time evolves, planetesimals are cleared from these resonant regions. Since there appears to be no difference in the early-time collision rate across simulations, we conclude that mean motion resonances do not drive the dynamics of collisional growth. They do, however, help to drive ejection of material on timescales of $10$ to $100 \; \mathrm{Myr}$. In \cjs{} configurations, planetesimals still populate the regions covered by the 3:1, 5:2, and 7:3 mean motion resonances ($2.5 \lesssim a \lesssim 3.2 \; \mathrm{AU}$). Over time, angular momentum exchange with Jupiter excites surviving planetesimals in this region onto hyperbolic orbits. This ejects them from the system.


%
%

\subsection{Disk Evolution}
\label{sec:sResults_ssEvolution}

\begin{figure*}
\includegraphics[scale=0.96]{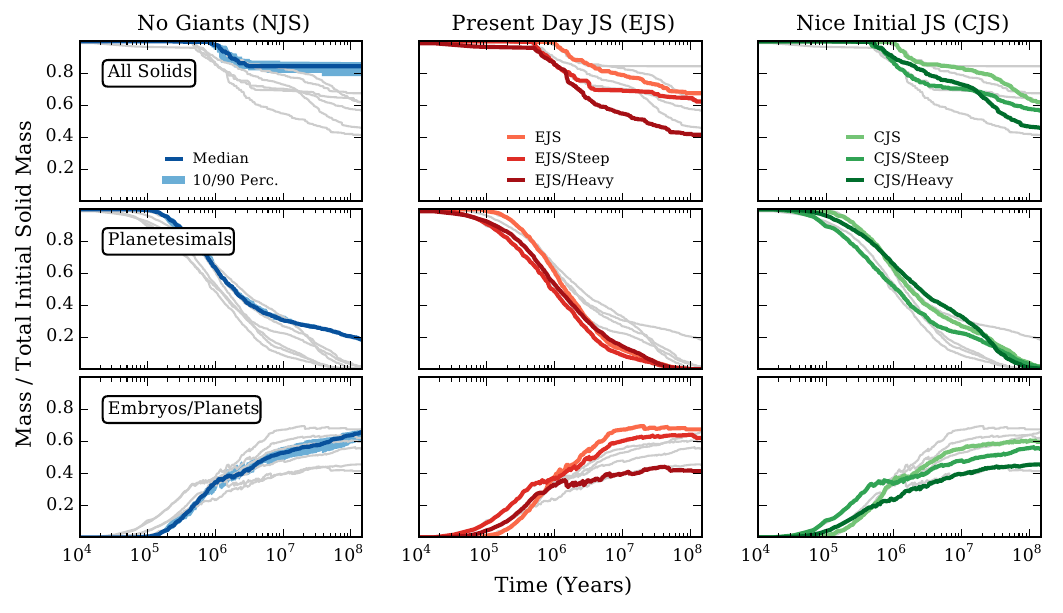}
\caption{Remaining solid mass (total, planetesimals, embryos) over time as a fraction of the total initial mass. We show the mass as a fraction of the initial total solid disk mass. In general, we show the median across twelve runs. For runs without Jupiter and Saturn, we also show the 10/90 percentile spread about the median. The spread is similar for all runs, so we do not indicate it separately for \ejs{} and \cjs{} runs. \textit{Rows (Top to Bottom):} Total solid disk mass, planetesimal mass, and total mass in embryos/planets as a function of time (rows, top to bottom). \textit{Columns (Left to Right):} Masses for runs without Jupiter and Saturn (\njs), runs with Jupiter and Saturn on eccentric orbits (\ejs), and runs with Jupiter and Saturn on circular orbits (\cjs). For \ejs{} and \cjs{} runs, we show results for three different initial planetesimal disks.}
\label{fig:fig03_mass_evolution}
\end{figure*}

\begin{figure}
\centering
\includegraphics[scale=0.76]{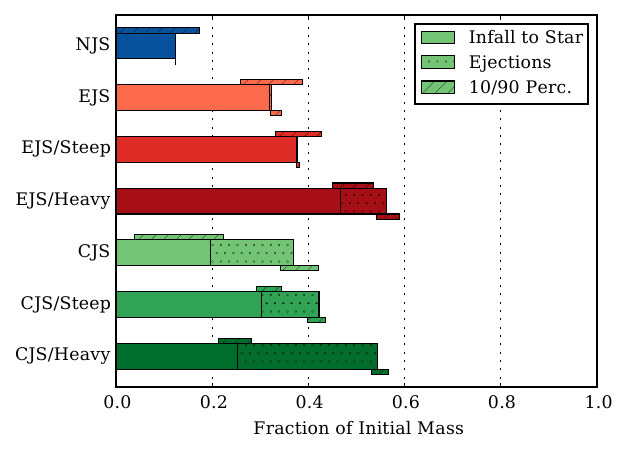}
\caption{Fraction of initial solid disk mass that falls onto the star (heliocentric distance $r < 0.2 \; \mathrm{AU}$) or is ejected from the system ($r > 20.0 \; \mathrm{AU}$). We show the median over twelve runs for each set of simulations as the main bars. The small bars below and above indicate the $10/90$ percentile spread of ejected and mass deposited onto the star, respectively. The $10/90$ percentile ranges for \njs{} runs are unreliable due to file corruption.}
\label{fig:fig04_ejection_infall}
\end{figure}

\renewcommand{\arraystretch}{1.3}
\begin{table}
\caption{Total disk mass and total mass locked up in embryos. We show median mass, offsets from median to $25/75$ percentile, and offsets from $25/75$ to $10/90$ percentile as sub-/superscripts.} 
\label{table:mass_evolution_01}
\begin{center}
\begin{tabular}{c c c c}
\toprule

  Set
& Time
& $M_\mathrm{Disk}$\textsuperscript{a,c}
& $M_\mathrm{Embryos}$\textsuperscript{b,c} \\

\midrule

\njs & $1 \; \mathrm{Myr}$ & $4.87^{+0.04(+0.04)}_{-0.05(-0.09)}$ & $1.73^{+0.07(+0.04)}_{-0.09(-0.15)}$ \\
& End & $4.22^{+0.09(+0.04)}_{-0.15(-0.16)}$ & $3.30^{+0.08(+0.05)}_{-0.13(-0.14)}$ \\
\midrule
\ejs & $1 \; \mathrm{Myr}$ & $4.91^{+0.07(+0.01)}_{-0.12(-0.05)}$ & $1.95^{+0.06(+0.03)}_{-0.12(-0.09)}$ \\
& End & $3.38^{+0.15(+0.06)}_{-0.27(-0.04)}$ & $3.38^{+0.13(+0.07)}_{-0.27(-0.04)}$ \\
\midrule
\cjs & $1 \; \mathrm{Myr}$ & $4.92^{+0.06(+0.01)}_{-0.06(-0.02)}$ & $1.83^{+0.05(+0.08)}_{-0.07(-0.04)}$ \\
& End & $3.09^{+0.06(+0.26)}_{-0.15(-0.26)}$ & $2.99^{+0.08(+0.24)}_{-0.10(-0.28)}$ \\
\midrule
\ejssteep & $1 \; \mathrm{Myr}$ & $4.26^{+0.08(+0.04)}_{-0.09(-0.04)}$ & $1.84^{+0.09(+0.07)}_{-0.08(-0.04)}$ \\
& End & $3.12^{+0.12(+0.10)}_{-0.21(-0.05)}$ & $3.11^{+0.12(+0.10)}_{-0.20(-0.05)}$ \\
\midrule
\ejsheavy & $1 \; \mathrm{Myr}$ & $8.51^{+0.18(+0.18)}_{-0.19(-0.03)}$ & $3.24^{+0.26(+0.14)}_{-0.07(-0.30)}$ \\
& End & $4.15^{+0.35(+0.14)}_{-0.27(-0.33)}$ & $4.15^{+0.34(+0.15)}_{-0.28(-0.34)}$ \\
\midrule
\cjssteep & $1 \; \mathrm{Myr}$ & $4.28^{+0.10(+0.06)}_{-0.09(-0.07)}$ & $1.70^{+0.13(+0.09)}_{-0.04(-0.07)}$ \\
& End & $2.83^{+0.13(+0.10)}_{-0.08(-0.11)}$ & $2.78^{+0.11(+0.11)}_{-0.14(-0.09)}$ \\
\midrule
\cjsheavy & $1 \; \mathrm{Myr}$ & $8.95^{+0.09(+0.04)}_{-0.15(-0.07)}$ & $2.34^{+0.23(+0.16)}_{-0.09(-0.02)}$ \\
& End & $4.57^{+0.15(+0.23)}_{-0.24(-0.35)}$ & $4.54^{+0.16(+0.20)}_{-0.26(-0.38)}$ \\

\bottomrule
\end{tabular}
\end{center}
\raggedright
\textsuperscript{a} Total mass in disk.
\textsuperscript{b} Total mass in embryos.
\textsuperscript{c} In Earth Masses.\\
\end{table}
\renewcommand{\arraystretch}{1.0}

\renewcommand{\arraystretch}{1.0}
\begin{table}
\caption{Percentile offsets from median: (i) total mass, (ii) total mass in embryos and (iii) total mass in planetesimals. The offsets are averaged (50 percentile) in time.}
\label{table:mass_evolution_02}
\begin{center}
\begin{tabular}{c c c c c}
\toprule

  Set
& Percentile
& $M_\mathrm{Disk}$\textsuperscript{a,d}
& $M_\mathrm{Embryos}$\textsuperscript{b,d}
& $M_\mathrm{Planetesimals}$\textsuperscript{c,d,e} \\

\midrule

\njs & $90$ & $+0.13$ & $+0.12$ & $+0.08$ \\
    & $75$ & $+0.10$ & $+0.10$ & $+0.03$ \\
    & $25$ & $-0.15$ & $-0.15$ & $-0.04$ \\
    & $10$ & $-0.31$ & $-0.27$ & $-0.06$ \\

\midrule

\ejs & $90$ & $+0.27$ & $+0.24$ & $+0.18$ \\
    & $75$ & $+0.15$ & $+0.13$ & $+0.06$ \\
    & $25$ & $-0.12$ & $-0.22$ & $-0.05$ \\
    & $10$ & $-0.32$ & $-0.32$ & $-0.08$ \\

\midrule

\cjs & $90$ & $+0.22$ & $+0.27$ & $+0.15$ \\
    & $75$ & $+0.10$ & $+0.08$ & $+0.11$ \\
    & $25$ & $-0.13$ & $-0.08$ & $-0.07$ \\
    & $10$ & $-0.27$ & $-0.22$ & $-0.15$ \\

\midrule

\ejssteep & $90$ & $+0.12$ & $+0.13$ & $+0.13$ \\
    & $75$ & $+0.09$ & $+0.09$ & $+0.07$ \\
    & $25$ & $-0.19$ & $-0.19$ & $-0.03$ \\
    & $10$ & $-0.30$ & $-0.32$ & $-0.04$ \\

\midrule

\ejsheavy & $90$ & $+0.51$ & $+0.50$ & $+0.11$ \\
    & $75$ & $+0.35$ & $+0.34$ & $+0.05$ \\
    & $25$ & $-0.26$ & $-0.26$ & $-0.07$ \\
    & $10$ & $-0.60$ & $-0.63$ & $-0.14$ \\

\midrule

\cjssteep & $90$ & $+0.20$ & $+0.16$ & $+0.10$ \\
    & $75$ & $+0.14$ & $+0.08$ & $+0.04$ \\
    & $25$ & $-0.10$ & $-0.15$ & $-0.06$ \\
    & $10$ & $-0.20$ & $-0.24$ & $-0.12$ \\

\midrule

\cjsheavy & $90$ & $+0.34$ & $+0.33$ & $+0.37$ \\
    & $75$ & $+0.13$ & $+0.23$ & $+0.23$ \\
    & $25$ & $-0.27$ & $-0.25$ & $-0.23$ \\
    & $10$ & $-0.61$ & $-0.53$ & $-0.35$ \\

\bottomrule
\end{tabular}
\end{center}
\raggedright
\textsuperscript{a} Total mass in disk.
\textsuperscript{b} Total mass in embryos.
\textsuperscript{c} Total mass in planetesimals.
\textsuperscript{d} In Earth Masses.
\textsuperscript{b} Only considers times $<32.9 \; \mathrm{Myr}$.\\
\end{table}
\renewcommand{\arraystretch}{1.0}

We now explore how the collisional evolution for a given set of simulations differs. Figure \ref{fig:fig03_mass_evolution} shows the total mass bound in (i) the disk, (ii) planetary embryos, and (iii) surviving planetesimals. As in \cite{2010Icar..207..517M}, a planetary embryo is an object with mass $M > M_\mathrm{Cut} = 3.3 \times 10^{26} \; \mathrm{g}$, which is about the mass of Mercury. Although embryos grow into planets, we do not label them separately. Objects with $M < M_\mathrm{Cut}$ are classified as planetesimals. For all simulation sets, we show the median mass per component. For the \njs{} runs, we also indicate the $10/90$ percentile spread about the median. In all simulation sets, the spreads are comparable in magnitude to the \njs{} case, so we omit them for the other sets to remove visual clutter. Tables \ref{table:mass_evolution_01} and \ref{table:mass_evolution_02} tabulate the ranges indicated in Figure \ref{fig:fig03_mass_evolution}. In Figure \ref{fig:fig04_ejection_infall}, we additionally indicate the median amount of material that is ejected from the system or falls onto the star along with the $10/90$ percentile range.

%
%

Initially, all simulation sets evolve identically. After $10^5$ years, the first planetary embryos become visible. It takes about four equal mass collisions to reach the cutoff mass, so the observed early-time collision rate of $\approx 0.03$ collisions per year per $2000$ particles agrees with this build-up. After $1 \; \mathrm{Myr}$, all simulations have assembled between $1.5$ and $2.0 \; \Mearth$ into embryos, but have overall lost very little of their total solid disk mass. Less than half the systems have lost more than $0.1 \; \Mearth$, although isolated systems in configurations without giant panets have lost as much as $0.55 \; \Mearth$. By the $3 \; \mathrm{Myr}$ mark, total mass loss across all runs has increased to between $0.54$ and $1.14 \; \Mearth$. The mass-loss is driven by hydrodynamic drag and type I migration, which by now have delivered particles to the inner edge of the disk, where we remove them from the simulation. In fact, the first particles already fall in around $2 \times 10^5$ years, but they do not carry significant mass. After $3 \; \mathrm{Myr}$, the gas disk is depleted by a factor of $20$ and becomes dynamically irrelevant for migration \citep{2002ApJ...565.1257T} and secular resonance sweeping \citep{2000AJ....119.1480N}, although it remains important for damping the eccentricies and inclinations of embryos \citep{2002Icar..157...43K,2004ApJ...602..388T}.\footnote{Following Eqns. (70) and (49) in \cite{2002ApJ...565.1257T} and \cite{2004ApJ...602..388T}, respectively, yields -- for a $1 \; \Mearth$ planet at $1 \; \mathrm{AU}$ in gas disk depleted by a factor of $10^4$ -- a migration timescale $\dot{a}/a \approx 10^9$ years, but a damping timescale of $\dot{e}/e \approx 10^7$ years, which is well-within the bounds for relevance found by \cite{2002Icar..157...43K}.} At this stage, all simulations host between $5$ and $18$ oligarchic embryos at semi-major axes $a \lesssim 2 \; \mathrm{AU}$.

During the first $3 \; \mathrm{Myr}$, the differences between sets are small, largely because gas drag acts as an equaliser that dampens eccentricity excitements from resonant interaction between planetesimals and Jupiter. At later times, we observe three marked differences. First, simulations with giant planets are much more efficient at assembling embryos. At $\sim 50 \; \mathrm{Myr}$, \ejs{} simulations have converted all low-mass particles. \cjs{} simulations are slower, and retain $\sim 0.08 \; \Mearth$ in the low-mass regime when the simulations terminate. The process is even slower in simulations without giant planets, which retain $\sim 1 \; \Mearth$ in low-mass objects at termination. Second, simulations with giant planets continue to lose mass after $3 \; \mathrm{Myr}$, eventually bringing the remaining disk mass down to $\sim 3.38 \; \Mearth$ (\ejs) and $\sim 3.09 \; \Mearth$ (\cjs). Third, there is a significant difference in how \ejs{} and \cjs{} simulations lose disk mass. For \ejs{}, all lost mass is lost onto the host star, while \cjs{} simulations eject about half the lost mass from the system. However, these processes are limited to late times.

The spread of tracked mass ranges is never uncomfortably large, such that the $10$ and $90$ percentiles are never more than $15$ per cent off the median. The exception to this is the total planetesimal mass, which suffers from small number statistics as planetesimals deplete at times $\gtrsim 100 \; \mathrm{Myr}$. Overall, the tight spread suggests that the range of evolutionary paths available to individual runs in a set is limited. However, we do observe three key trends in statistical spread with simulation sets. First, the spread is consistently smallest in masses below the cutoff mass. Second, the spread across runs in total disk mass is larger in simulations that include giant planets. Third, the probability of larger excursions is higher in the \ejs{} and \cjs{} simulations. For \ejs{} sets, the $10/90$ percentiles tend to be a factor $2$ further off the median than the $25/75$ percentiles. In \cjs{} sets, the difference grows and exceeds factors of $3$ for the $90$ percentile in mass above the cutoff.

%
%

At small semi-major axis, relative velocities between planetesimals are higher, leading to shorter timescales of collision and thus growth. Runs with steeper initial surface density profiles therefore start assembling embryos already after $\sim 2 \times 10^4$ years. As a consequence, more embryos are driven close to the star by type I migration, where they are removed from the simulation. This effectively stalls the conversion of planetesimals into embryos between $5 \times 10^5$ and $10^6$ years. Although the conversion process picks up again, initially steep disks still tend to host $\sim 8$ per cent less total mass in terrestrial planets than the reference disks. Apart from the short stall, the fraction of mass in embryos grows continuously until $\sim 10 \; \mathrm{Myr}$, whereafter it slows down significantly. This is irrespective of the orbit of the giant planets, although conversion of mass into embryos proceeds slightly faster if giant planets are on eccentric orbits.

In massive disks, conversion of planetesimals into embryos begins earlier than in the reference run (but later than in runs with steep initial surface density profiles). This is likely a numerical artefact related to the doubling of the initial planetesimal mass.\footnote{In massive disks, only three instead of four equal mass collisions are required to reach the cutoff mass for classification as embryo. To avoid such numerical effects, future simulations that vary the total disk mass should keep a constant mass resolution.} More robustly, we find that the fraction of mass in embryos either (i) stops increasing significantly after $\sim 1 \; \mathrm{Myr}$ (eccentric giant planets) or (ii) proceeds slower overall (circular giant planets) for initially massive disks. However, in absolute terms, it does settle at a higher level ($6 \; \Mearth$ vs $\sim 3 \; \Mearth$ for less massive disks).

Despite the slower rate of embryo assembly and smaller total mass fraction of embryos, both steep and massive disks lose total mass earlier and more efficiently than the reference profiles. For the reference disks, mass loss only sets in after $\sim 1 \; \mathrm{Myr}$; almost $0.5 \; \mathrm{Myr}$ later than when steep and massive disks begin loosing material. Most extremely, systems hosting massive disks with eccentric giant planets begin loosing mass almost immediately at the $2 \times 10^4 \; \mathrm{yr}$ mark. These systems must be dynamically more active than the reference profiles, such that more orbits can deliver mass onto the star, out of the system, or onto the giants.

In fact, no single planetesimal or embryo collides with Saturn, and we record only three to four planetesimals (corresponding to $0.01$ to $0.05 \; \Mearth$) impacting on Jupiter per run, but only in runs with massive planetesimal disks. There are occasional collisions with Jupiter in other runs, but these are isolated events and do not occur in all runs of a given configuration. This is not particularly surprising as Jupiter and Saturn have larger escape velocities than the host star (evaluated at their location). In other words, particle from the inner systems that have been thrown onto orbits crossing Jupiter or Saturn tend to be too fast to be caught by either \citep{2008ApJ...686..621F,2010ApJ...711..772R}. Most of the mass loss is thus onto the star or through ejection from the system and the mechanics vary depending on the configuration, see Figure \ref{fig:fig04_ejection_infall}. In systems without giant planets and with giant planets on eccentric orbits, almost all mass is lost onto the star, although massive disks in \ejs{} configurations eject $\sim 10$ per cent of the lost mass. For configurations of giants on circular orbits, between $20$ and $50$ per cent of the lost solid mass is ejected. For both giant planet configurations, initially massive disks appear to eject the largest fraction of their initial mass.

Similarly to the reference runs, the spread of the masses remains well-bound. Except for outliers from small number statistics in the planetesimal mass at late times, the $10$ and $90$ percentiles of the masses remain within $15$ per cent off the median. For eccentric giant planets, steep initial disks have a smaller and massive initial disks a larger spread than the reference case. For giants on circular orbits, steep profiles have comparable spreads to the initial profile. Massive planetesimals disks in \cjs{} configurations show no definite trends either. For example, the $90$ percentile mass in embryos in massive disks has a smaller offset from the median than the reference profile ($7.2$ vs. $9.2$ per cent) while the $75$ percent has a larger offset ($5.17$ vs. $2.70$ per cent).


\subsection{Mass Distributions \& Connection to Initial Conditions}
\label{sec:sResults_ssArch}

\begin{figure*}
\includegraphics[scale=0.96]{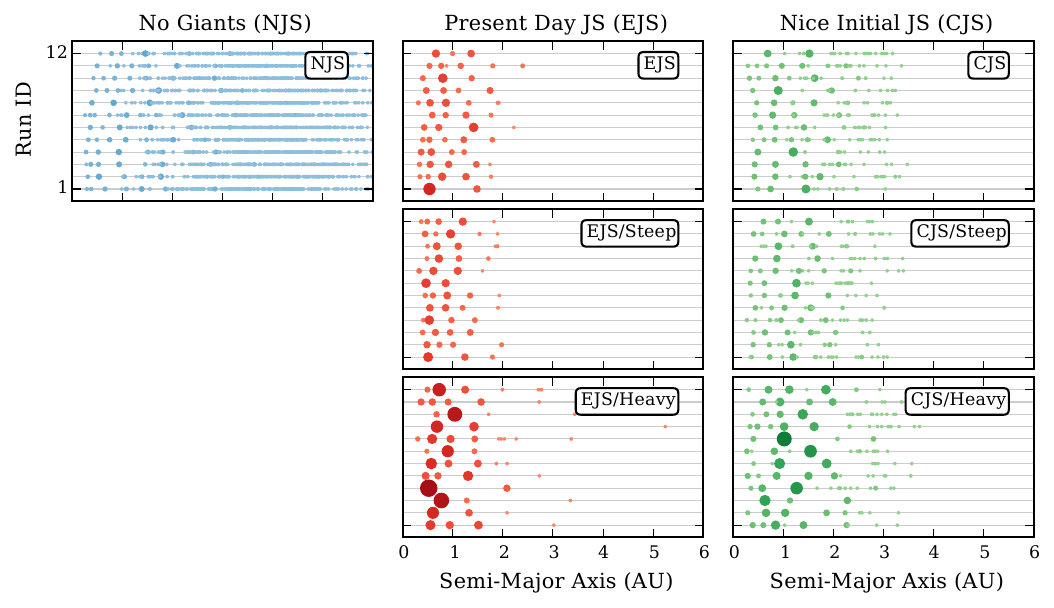}
\caption{Distribution of planets/planetesimals with semi-major axis for all $7 \times $12 initially identical runs after $147.84 \; \mathrm{Myr}$. Each line indicates a separate run. Darker colours and bigger circles indicte more massive particles. \textit{Columns (Left to Right)}: Runs without Jupiter and Saturn, Jupiter and Saturn on their present-day eccentric orbits, Jupiter and Saturn on circular orbits. \textit{Rows (Top to Bottom)}: Reference initial planetesimal disk, steep disk, more massive initial disk.}
\label{fig:fig05_system_architecture}
\end{figure*}

\begin{figure*}
\includegraphics[scale=0.96]{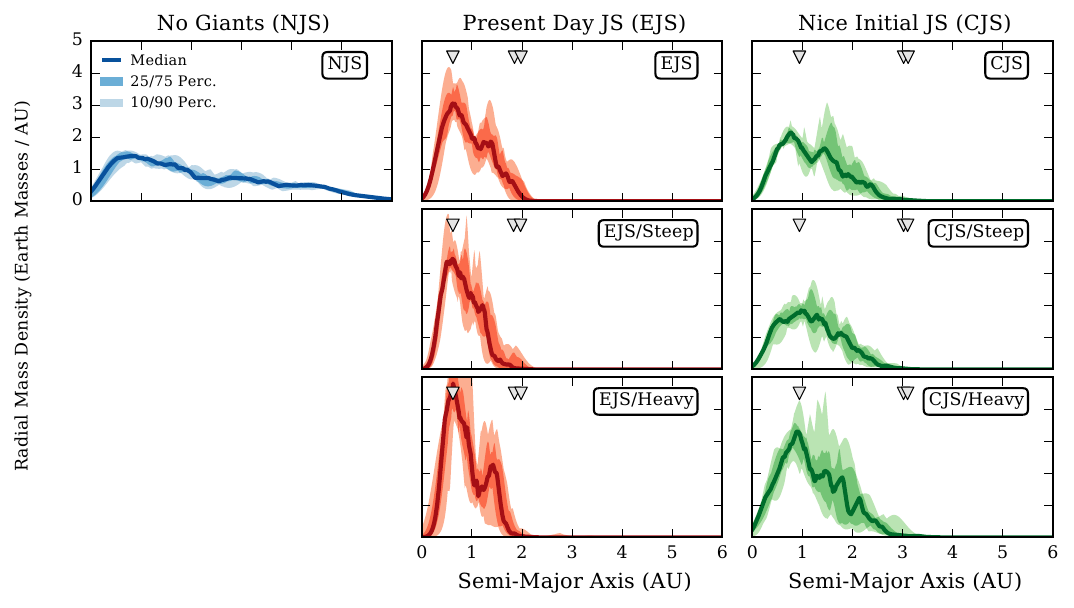}
\caption{Median (lines) and statistical spreads (shading) of the radial mass distribution each set of $7\times 12$ runs shown in Figure \ref{fig:fig05_system_architecture}. The distributions are obtained as kernel density estimates of the mass-weighted particle distribution $\mathrm{d}M/\mathrm{d}a$ along the semi-major axis $a$. Black lines indicate the median. The shadings indicate $25/75$ and $10/90$ percentile ranges. Tiangles indicate (from left to right) the locations of the $\nu_5$, $\nu_6$, and $\nu_{16}$ secular resonances for runs with giant planets. \textit{Columns (Left to Right)}: Runs without Jupiter and Saturn, Jupiter and Saturn on their present-day eccentric orbits, Jupiter and Saturn on circular orbits. \textit{Rows (Top to Bottom)}: Reference initial planetesimal disk, steep disk, massive initial disk.}
\label{fig:mass_distribution_in_semi_major_axis}
\end{figure*}

\begin{figure*}
\centering
\includegraphics[scale=0.96]{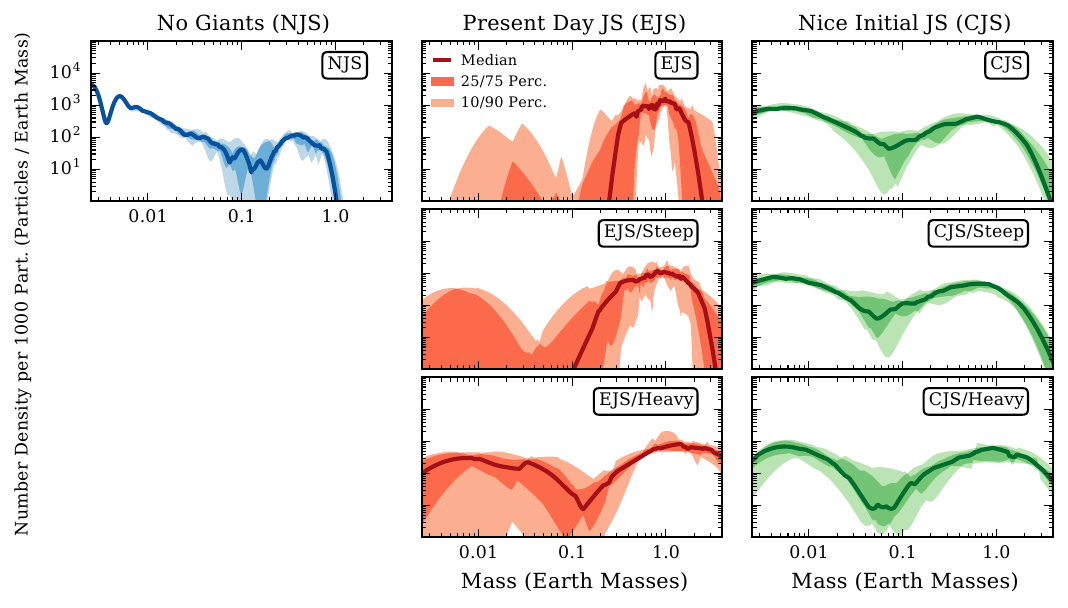}
\caption{Median (lines) and statistical spreads (shading) of the mass-frequency distribution (i.e., the number of particles per mass bin -- $\mathrm{d}N/\mathrm{d}M$) for our simulation sets. The distributions are computed as kernel density estimates. \textit{Columns (Left to Right)}: Runs without Jupiter and Saturn, Jupiter and Saturn on their present-day eccentric orbits, Jupiter and Saturn on circular orbits. \textit{Rows (Top to Bottom)}: Reference initial planetesimal disk, steep disk, more massive initial disk.}
\label{fig:mass_function_differential}
\end{figure*}

In terms of aggregate mass bound up in planetesimals, embryos, and the solid disk as a whole, runs within sets of initially identical conditions have variance below the $15$ per cent level. But what about the radial mass distributions and mass-frequency distributions at the end of the simulations? Do the systems look architecturally similar? Is the mass distributed similarly? What is link between the initial and final mass distributions?

Figure \ref{fig:fig05_system_architecture} illustrates the diverse architectures of systems we have generated after $147.84 \; \mathrm{Myr}$. Considering the twelve runs within sets, find we the arrangement of final planets to vary substantially. Nevertheless, the impact of adding or adjusting the configuration of the giant planets has an even stronger impact that can easily be discerned visually over the different architectures generated by the same  configuration. On the other hand, runs with different initial surface density profiles appear difficult to visually discern from the stochastic variations in architecture. For massive disks, this is easier because they tend to host massive terrestrial planets.

To be more quantitative, we analyse the radial mass distributions (mass per semi-major axis) and mass-frequency distributions (particle counts per mass bin; also called the mass function in cosmology and galaxy formation) for various giant planets configurations and initial planetesimal distributions. In both cases, we use non-parametric kernel density estimates (KDE) to derive the distribution functions for all runs in a set. They are combined statistically in Figures \ref{fig:mass_distribution_in_semi_major_axis} and \ref{fig:mass_function_differential}. We also compute the final surface density profiles by grinding up the terrestrial planets, distributing their mass over annuli of suitable widths, and fitting power-laws.\footnote{Profiles of surface density and radial mass distribution are related, but different methods of describing how mass is distributed. The radial mass distribution is a more general method relying on a non-parametric fit (generated by dropping Gaussians of adaptive width, summing them up, and then normalising the integral) whereas the surface density requires more manual intervention. Although a more general method is generally preferred, much of the literature is based on surface densities.} In Figure \ref{fig:fig08_memory}, we compare the range of power-law exponents for the planetary systems to the initial distribution of planetesimals.

\subsubsection{System Architectures \& Mass Distributions}

%
%

Simulations without giant planets distribute mass most evenly with semi-major axis, while \cjs{} and \ejs{} simulations concentrate progressively more mass at smaller semi-major axes. The \ejs{} systems tend to be truncated around $2 \; \mathrm{AU}$. In \cjs{} systems, terrestrial planets populate a region out to $3.5 \; \mathrm{AU}$, but most mass is concentrated within $2.2 \; \mathrm{AU}$. In simulations with giant planets, there appear to be regions that are preferably populated by terrestrial planets. Most notably, these are at $0.8$, $1.2$ (\ejs) as well as at $0.9$, $1.5$, and $2.2 \mathrm{AU}$ (\cjs), although the last peak is arguably weak. However, we observe significant variations in the distribution of mass and planets with semi-major axis within sets. These are smallest in systems without giant planets and largest in systems with giants on circular orbits. For example, the $90$ percentile masses are (by median) $25$, $50$, and $69$ per cent off the median mass (in order, \njs, \ejs, \cjs).

The lack of giant planets severely stunts growth of terrestrial planets, and these systems retain larger numbers of planetesimals at their origin mass. In contrast, \ejs{} configurations are the most efficient at assembling terrestrial planets with masses on the order of an Earth mass, while \cjs{} runs populate a middle ground. Irrespective of presence and configuration of giant planets, there is significant variance in the mass-frequency distribution. For \njs{} runs, the scatter is most significant at intermediate mass ranges $0.03$ to $0.3 \; \Mearth$, with the $25/75$ percentiles off by a factor of about two from the median. The same holds for \cjs{} configurations. For \ejs{} configurations, the mass-frequency distributions are more difficult to interpret because almost no planetesimals remain. Focussing on planetary masses ($0.5$ to $1.5 \; \mathrm{AU}$), we find a larger spread across \ejs{} runs than for \cjs{} runs ($10$ to $20$ per cent vs. factors of two).

%
%

Initialising the planetesimals to follow a steeper surface density profile has a surprisingly small effect. For the \cjs{} case, we find no obvious discernible trends, although we do note two changes for \ejs{} configurations. Firstly, more mass is concentrated at smaller semi-major axis, which is a consequence of more mass being initially distributed here. This is only significant in \ejs{} configurations because the semi-major axis range populated by terrestrial planets is tighter here. Secondly, the mass range covered by planets widens a bit and we find that more \ejs{} runs retain some planetesimals in initially steep disks. This still holds for less than half of the runs, which is still more than in the reference profile.

Increasing the initial disk mass has a more drastic effect on the distributions. Across the entire semi-major axis range populated by the terrestrial planets, both \cjs{} and \ejs{} configurations record a higher median mass as well as a larger scatter. Massive initial disks also generate a wider range of planet masses (both at the lower and upper mass end), which essentially flattens out the mass-frequency distribution, although the dip around $0.1 \; \Mearth$ in \cjs{} configurations remains and exhibits a large scatter. For \ejs{} runs, we find more (but still not all) simulations to retain planetesimals, which puts power into the lower end of the mass-frequency distribution. Nevertheless, large spreads remain here as the number of surviving planetesimals per run is still in the single digits.

%
%

Before closing, we wish to comment on the robustness of using kernel density estimates (KDEs) to characterise radial mass and mass-frequency distributions. Fitting a KDE amounts to attempting to estimate the underlying distribution of planetesimals that is sampled by the simulation particles. For each sample (particle), the algorithm drops a Gaussian kernel of a width depending on the spacing of samples. The KDE is then sum of kernels. For \njs{} runs, this works well because the distribution per run is fit to between $150$ and $200$ samples. For \cjs{} and \ejs{} configurations, the number of available samples drops to between $10$ and $20$ (\cjs), respectively $5$ to $10$ (\ejs). The algorithm compensates by dropping much wider Gaussian kernels. Stacking runs and computing percentiles over these kernels leads to wide regions with jagged edges that are especially apparent in the mass-frequency distributions of \ejs{} and \ejssteep{} runs. We thus caution from attributing too much meaning to these regions because they are very sparsely sampled.

\subsubsection{A Memory of Initial Conditions?}

\begin{figure}
\centering
\includegraphics[scale=0.82]{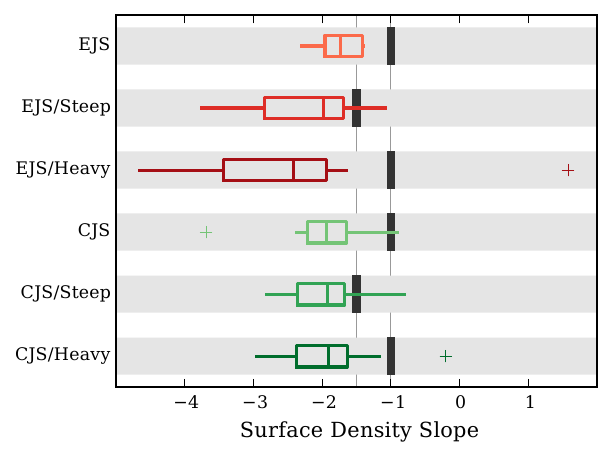}
\caption{Power-law slopes ($\beta$, where $\Sigma_\mathrm{Planets} \varpropto a^{\beta}$) for the projected surface densities $\Sigma_\mathrm{Planets}$ at semi-major axis $a$ obtained by grinding up the final terrestrial planets. Smaller values of $\beta$ indicate steeper profiles. Boxes indicate the $25/75$ percentile range, their centreline the median, whiskers extend to $1.5$ times the interquartile range, and crosses indicate outliers. We also show the slope of the initial planetesimal distribution with bold vertical lines.}
\label{fig:fig08_memory}
\end{figure}

Frequently, attempts at reproducing the formation history of both the solar system as well as exoplanetary systems are based around the concept of a Minimum Mass (Extra-)Solar Nebula. Here, the presently observed distribution of planets is ground up, distributed over their assumed feeding zones, and used to reconstruct a radial profile of the solid components that feed into the planets. By initialising simulations to follow such a profile, it is implicitly assumed that such a profile persists as planets form. For reference, \cite{1977Ap&SS..51..153W} and \cite{1981PThPS..70...35H} determine the Minimum Mass Solar Nebular to follow a surface density profile of $\Sigma_\mathrm{Planets} \varpropto r^{\; \beta}$ with $\beta \sim -1.5$. For extrasolar systems, \cite{2004ApJ...612.1147K} and \cite{2013MNRAS.431.3444C} report $\beta \sim -1.6$ to $\beta \sim -2.4$. More recently, \cite{2014MNRAS.440L..11R} report an even wider range of profiles that can be fit through observed systems ($\beta \sim -3.2$ to $\beta \sim 0.5$) and argue this indicates that there is no (or no tight range of) underlying initial profiles. We now show the ranges of final surface density profiles that can be obtained from a given initial distribution to determine if the initial profile indeed persists. We will find that -- even when starting with identical initial conditions -- there is large scatter of generated profiles, such that the naive mapping of an initial to a final profile does not make sense.

We now compute how the surface density of our terrestrial planets $\Sigma_\mathrm{Planets}$ varies with semi-major axis $a$ at the end of each of our runs. We grind up each planet $k$ over an annulus with boundaries determined by the geometric mean to the next planet on either side \citep{1977Ap&SS..51..153W,2004ApJ...612.1147K} to recover $\Sigma_{\mathrm{Planet},k}$ and then fit a power-law of the form $\Sigma_\mathrm{Planets} \varpropto a^{\, \beta}$. We perform the least-squares fit in log-space and consider a coefficient of determination $r^2 > 0.6$ to indicate a sufficient enough fit to allow such a power-law description. Figure \ref{fig:fig08_memory} summarises the statistical spread of the obtained power-law exponents. In each of simulation set, at most three runs have $r^2 \leq 0.6$, which we exclude from the analysis. Note that we do not include \njs{} runs because they are dominated by a large number of small embryos or large planetesimals where the concept of ``grinding up planets'' does not make sense.


In $\sim 85$ per cent of the final systems, the surface density profiles are steeper than the initial profiles, with rare occurrences of a shallower final profile existing in all initial planetesimal distributions and giant planet configurations. For the \ejs{} runs, the final profiles are always steeper than the initial ones. The median steepening is largest for \ejsheavy{} and \cjs{} ($142$ and $94$ per cent, respectively) and smallest for \ejssteep{} and \cjssteep{} runs ($32$ and $29$ per cent, respectively).

All in all, the results are sobering. Although a power-law is good fit to over $75$ per cent of the final distributions, there is a huge scatter in fitted exponents across runs within a set. Even in the most robust case (\ejs), the difference between the shallowest and steepest profiles exponents are about $1$ with a $25$ per cent scatter between the $10/90$ percentiles and the median. For all other configurations, the average spread exceeds $40$ per cent, but we note that the $10$ percentiles tend to be within $30$ percent of the median, while the $90$ percentile is usually $60$ to $90$ per cent off the median. The largest absolute difference between $10$ and $90$ percentile exponents is recorded for \ejsheavy{} with a difference of $2.23$. Clearly, generating a simple mapping from initial to a final distribution is out of the question.

Interestingly enough, when \cite{2005ApJ...632..670R} characterised their difference between initial and final surface density profiles, they found little change in the slope of the final power-law, except for profiles that had substantial amount of material in the vicinity a giant planet. However, they started from a distribution of embryos which did not grow from planetesimals in a gaseous disk. As such, their planets were relegated to forming locally based on the initial distribution. We therefore conclude that a mapping initial to final profiles is only possible if the initial profile is considered at a stage when embryos have already formed, gas has dissipated, and there is no immediate influence of giant planets. Conversely, is it not possible to establish a link to the initial distribution of planetesimals in a gas disk by observing present-day systems.


\subsection{Stability \& Orbital Spacing }

On timescales of a few $100 \; \mathrm{Myr}$, stability for two orbiting planets against close encounters requires their orbital separation $\Delta a$ to exceed $\Delta a > 10 \; \Rhill$, where $\Rhill = ((m_1 + m_2)/(3 \Msun))^{1/3} \, (a_1+a_2)/2$ is the mutual Hill radius of two neighbouring planets \citep{1993Icar..106..247G,1996Icar..119..261C}. The threshold increases to $\Delta a \gtrsim 20$ for systems with $10$ to $20$ planets. By this criterion, we find about $90$ per cent of all terrestrial planets to be on stable orbits although this number fluctuates with giant planet configuration and initial planetesimal disk profile. The configurations \ejssteep{} and \ejsheavy{} have no planet pairs with spacings $\Delta a < 20 \; \Rhill$. We therefore expect none of these runs to eject (or collide) any planets with a few hundred $\mathrm{Myr}$ after our simulations terminate. Of the remaining configurations, the \cjsheavy{} simulations have the fewest pairs spaced $\Delta a < 20 \; \Rhill$, so that we expect at most one planet for every four runs to collide or be ejected. The systems in \ejs, \cjs, and \cjssteep{} even admit a few extremely tightly planets with $\Delta a < 3 \; \Rhill$. Over twelve runs, we expect one planet to be ejected per every two to three runs. Finally, \njs{} configurations without giant planets host a large number of such tight separations. Here, we expect at least one to two planets per run to collide or be ejected within a few hundred $\mathrm{Myr}$ after our runs terminate.


\cite{1999Icar..139..328Y} improve on this stability analysis by considering the influence of eccentric (and inclined) orbits. They use the scaled eccentricity $\tilde{e} = e/h$ (with $h=\Rhill/a$) in systems of $10$ protoplanets with mutual spacing $4 \leq \Rhill \leq 10$ and $0 \leq \tilde{e} \leq 4$ to find that eccentric orbits shorten the time it takes a set of protoplanets to become dynamical unstable by up to two orders of magnitude. Unfortunately, it is unclear to what degree their analysis can be applied to our simulations as they cover an entirely different dynamic regime. Except for \njs{} runs, we find systems that host $4$ to $8$ planets with spacings  $20 \lesssim \Delta a / \Rhill \lesssim 55$ and eccentricities $3 \lesssim \tilde{e} \lesssim 30$ (bounds are $10/90$ percentiles). Attempting a simple extrapolation of their fits, we find a required spacing of $40 \; \Rhill$ for planets excited to $\tilde{e} \sim 8$ (the median for our runs) to remain stable for a hundred Myr. For runs with giant planets, this is fulfilled for over $90$ per cent of all cases which suggests stability to be in line with the analysis based on \cite{1996Icar..119..261C} above. For \njs{} runs, only $25$ per cent of planets have such a large spacing, again emphasizing that we expect significant dynamical evolution in the systems after our simulations terminate. At the $90$ percentile level, our simulations exhibit values between $\tilde{e} \sim 14$ and $\tilde{e} \sim 36$, which requires separations of $\sim 80 \; \Rhill$ and $\sim 160 \; \Rhill$, which is not fulfilled by all our planets at such eccentricities. Overall, we conclude that the \cite{1999Icar..139..328Y} criteria are stricter than those of \cite{1996Icar..119..261C} and that their direct application would suggest a larger fraction of planets to not survive the next few hundred Myr in our simulations. However, the straightforward application of their criterion is questionable as our systems populate a different dynamical regime.

It remains now to characterise what sets the spacing of terrestrial planets. In particular, the double peaked structures of the radial mass distribution for some of our simulations (\ejs, \cjs, and \ejssteep) suggests preferred orbital spacings, which may be indicative of chains of mean motion resonances. However, out of a total of $35'554$ planet pairs, we find that only $13$ ($0.036$ per cent) are no further than $0.001 \; \mathrm{AU}$ away from a the semi-major axis ratio required for any of the mean motion resonances considered (3:1, 5:2, 7:3, 2:1, and 3:2). Even if we are willing to relax the distance to the resonance to $0.01 \; \mathrm{AU}$, only $117$ ($0.33$ per cent) of planet pairs are in resonance. We conclude that resonances between the formed terrestrial planet are rare and do not drive the double peaked radial mass distribution seen in some configurations.


\subsection{Final Orbital Parameters}
\label{ssec:orbits}

\begin{figure*}
\includegraphics[scale=0.96]{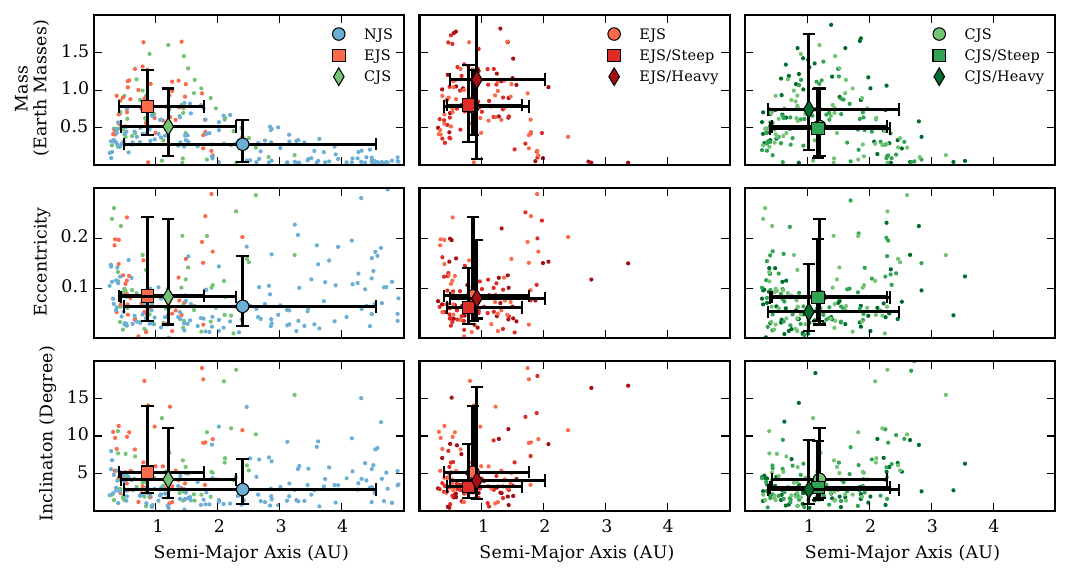}
\caption{Final distribution of (rows, from top to bottom rows) mass, eccentricity, and inclination with semi-major axis of all terrestrial planets in the simulations. Small markers indicate the formed terrestrial planets (we exclude remaining planetesimals). The large overlaid markers indicate the median and $10/90$ percentile range. Note that there exist points outside the view of the Figure, although their fraction of the total is at or below the 10 percent level. We chose a zoomed view to focus on the differences in median and percentile spreads. \textit{Columns (Left to Right)}: Reference initial planetesimal disk with different giant planet configurations, Jupiter and Saturn on present-day eccentric orbits for different planetesimal disks, Jupiter and Saturn on circular orbits for different planetesimal disks.}
\label{fig:fig09_orbits}
\end{figure*}

\renewcommand{\arraystretch}{1.3}
\begin{table}
\caption{Median terrestrial planet mass $M$, semi-major axis $a$, eccentricity $e$, inclination $i$ across all $12$ simulations of a set. $N$ is the median number of planets per simulation in a set. Sub-/superscripts are $10/90$ percentiles.}
\label{table:final_planets}
\begin{center}
\begin{tabular}{c c c c c c}
\toprule

Set & $M$ ($\Mearth$) & $a$ ($\mathrm{AU}$) & $e$ & $i$ ($\mathrm{Degree}$) & $N$ \\
\midrule
\njs & $0.28_{\;0.04}^{\;0.61}$ & $2.40_{\;0.49}^{\;4.55}$ & $0.06_{\;0.02}^{\;0.16}$ & $2.89_{\;1.01}^{\;6.92}$ & $12_{\;10}^{\;12}$ \\
\midrule
\ejs & $0.78_{\;0.40}^{\;1.27}$ & $0.86_{\;0.40}^{\;1.77}$ & $0.08_{\;0.03}^{\;0.24}$ & $5.18_{\;2.42}^{\;13.93}$ & $4_{\;3}^{\;5}$ \\
\ejssteep & $0.80_{\;0.32}^{\;1.33}$ & $0.79_{\;0.45}^{\;1.66}$ & $0.06_{\;0.03}^{\;0.14}$ & $3.31_{\;1.74}^{\;9.00}$ & $4_{\;3}^{\;4}$ \\
\ejsheavy & $1.14_{\;0.08}^{\;2.45}$ & $0.93_{\;0.49}^{\;2.02}$ & $0.08_{\;0.04}^{\;0.20}$ & $4.06_{\;1.68}^{\;16.48}$ & $3_{\;2}^{\;5}$ \\
\midrule
\cjs & $0.52_{\;0.12}^{\;1.02}$ & $1.20_{\;0.44}^{\;2.29}$ & $0.08_{\;0.03}^{\;0.24}$ & $4.23_{\;1.76}^{\;11.03}$ & $5_{\;4}^{\;6}$ \\
\cjssteep & $0.49_{\;0.10}^{\;1.01}$ & $1.17_{\;0.41}^{\;2.33}$ & $0.08_{\;0.03}^{\;0.20}$ & $3.14_{\;1.48}^{\;9.37}$ & $5_{\;4}^{\;6}$ \\
\cjsheavy & $0.75_{\;0.20}^{\;1.74}$ & $1.03_{\;0.38}^{\;2.48}$ & $0.05_{\;0.01}^{\;0.15}$ & $2.84_{\;0.95}^{\;9.51}$ & $5_{\;3}^{\;6}$ \\
\midrule
SS & $0.46_{\;0.07}^{\;0.94}$ & $0.86_{\;0.49}^{\;1.37}$ & $0.06_{\;0.01}^{\;0.17}$ & $1.93_{\;1.61}^{\;5.10}$ & $4_{\;4}^{\;4}$ \\

\bottomrule
\end{tabular}
\end{center}
\end{table}
\renewcommand{\arraystretch}{1.0}

Numerically speaking, all particle distributions that are evolved from the same initial conditions are equally valid configurations of the planetary system. We thus argue that (at present) we have no means of determining the \textit{actual} configuration of a simulated planetary system. This leaves us with two choices of analysis. We could (i) try to determine the most likely actual configuration or (ii) explore the range of permitted configurations. If we do the former, we may be tempted to declare the median of all configurations to be the most likely true configuration. This is by no means well-justified, and would require a large number of simulation runs to escape the confines of small number statistics. Given this complication, we settle for a survey of possible configurations.

We determine the range of the valid configurations at the end of the simulations by stacking data from all runs in a given set. We restrict our analysis to particles $M > M_\mathrm{Cut}$, but make no distinction between planets or planetary embryos. Figure \ref{fig:fig09_orbits} shows the raw data as well as median and $10/90$ percentile ranges of the semi-major axis $a$, mass $M$, orbital eccentricity $e$, and orbital inclination $i$. Results are also tabulated in Table \ref{table:final_planets} where we list the number of planets per system. We include the solar system in Table \ref{table:final_planets} purely for reference and stress that none of our simulations were set up with the explicit intent of reproducing the solar system.

%
%

The presence and configuration of giant planets has the largest effect on the final configuration of terrestrial planets. In the absence of giant planets, we find $10$ to $12$ terrestrial planets per system. Of these, $90$ per cent have mass $M < 0.6 \; \Mearth$, and half have $M < 0.3 \; \Mearth$. By number, they are spread evenly across a wide range of semi-major axes with more massive planets closer to the host star. If giant planets are present, the number of terrestrial planets per systems drops to $4$ or $5$. They are more massive with $50$ per cent having $M > 0.5$ to $0.8 \; \Mearth$, depending on the giant planet configuration. Most ($90$ per cent) are below $1.2 \; \Mearth$, but we find isolated $1.6 \; \Mearth$ planets. They also cover a much narrower range of semi-major axes and are closer to the host star.

Systems with Jupiter and Saturn on eccentric orbits (\ejs) tend to have (by median) $50$ per cent more massive planets than systems with Jupiter and Saturn on circular orbits (\cjs). They place terrestrial planets at smaller semi-maxis axis. Note that these are broad trends. Both configurations can produce terrestrial planets $M > 1 \; \Mearth$, although \ejs{} simulations are more likely to do so. Simulations with eccentric giants planets also tend to have one or two fewer planets per system. Except for a few outliers, all planets are in orbits with low inclination and eccentricity. Overall, $50$ per cent of all planets are at inclinations $i < 5^\circ$ and eccentricities $e < 0.1$. More inclined and eccentric orbits are restricted to simulations with giant planets, where $10$ per cent of planets can reach $i > 15^\circ$ or $e > 0.25$.

These observations are consistent with those of \cite{2003AJ....125.2692L}, who report systems hosting embryos on more eccentric orbits to result in fewer, more massive, and closer-in terrestrial planets. Excitation of planetesimal (and ultimately terrestrial planet) eccentricity increases in the presence of giant planets, especially when those are on eccentric orbits. \cite{2004Icar..168....1R} find similar correlations and also note that the total mass of terrestrial planets tends to be slightly lower in systems with more eccentric Jupiters. We observe the opposite (cf. Figure \ref{fig:fig03_mass_evolution}), but point out their use of a different initial surface density profile. While the baseline simulations in \cite{2007ApJ...669..606R} do not host giant planets, they have performed some tests on their influence. Overall, they find the effect of adding a Jupiter-mass planet to be very small (the mean terrestrial planet mass increases by $10$ per cent). However, their runs are initialised at a stage when planetary embryos have formed and the gaseous disk has dissipated.

%
%

For a given giant planet configuration, changing the slope of the initial planetesimal mass distribution has a much smaller effect than adjusting the mass of planetesimal disk. In particular, changing the reference profile to a steeper profile changes the median mass and semi-major axis of the formed terrestrial planets by less than $10$ per cent for both giant planet configurations. Additionally, the width of the range in mass and semi-major axis populated by $80$ per cent of terrestrial planets also remains similar. Orbital inclinations decrease by $26$ (\cjs{} to \cjssteep) to $36$ (\ejs{} to \ejssteep) per cent. Eccentricities are only visibly stunted for \ejs{} runs with the median eccentricity for \ejssteep{} about by 25 per cent below that of \ejs{} runs.

Previously, \cite{2005ApJ...632..670R} also modelled the effect of varying the initial surface density profiles. They noted that steepening the profile increases the (mean) mass of the most massive planet, accelerates their formation, increases mean number of plants in a given system, and leads to more planets at smaller semi-major axes. They did not observe a conclusive trend for the eccentricity of the final planets. Some of these trend are in line with ours (more massive planets at smaller semi-major axes) while others are not (number of planets per system) with an additional dependency on our particular configuration of giant planets. However, besides similar initial surface density profiles, they initialise simulations in a distinctly different way. Their runs launch with fully formed embryos (no gas and no planetesimals) and have a single giant planet on a circular orbit. In our runs, sweeping resonances drive the final architecture of our systems. This mechanism is absent in their runs and their final architecture is driven solely by planet-planet scattering and interaction with the giant at mean motion resonances.

Predictably, heavier planetesimal disks produce more massive terrestrial planets. The most massive planets are $3.71 \; \Mearth$ (\ejs) and $3.05 \; \Mearth$ (\cjs). At the $90$ per cent level, terrestrial planets in initially massive disks exceed those in the reference by $93$ ($70$) per cent in \ejs{} (\cjs) configurations. The situation is less clean-cut in in semi-major axis, eccentricity, and inclination. For systems with eccentric giant planets, there is weak trend for terrestrial planets to end up at larger semi-major axis, although though median only differs by $8$ per cent. For systems with giants on circular orbits, the trend is reversed such that the terrestrial planets tend to form $15$ per cent closer to the stars. Dynamically, planets tend to be less excited in initially massive disks. This holds especially in \cjs{} configurations where eccentricities and inclinations decrease by $35$ per cent. In \ejs{} runs, the effect is weaker. Here, eccentricities are approximately the same. Although inclinations are reduced by $20$ per cent for most planets, they in fact increase for at least a tenth of all planets.

Although starting at from entirely different initial conditions, both \cite{2006ApJ...642.1131K} and \cite{2007ApJ...669..606R} also find the planet mass to scale with the solid disk mass. However, \cite{2006ApJ...642.1131K} has very different initial conditions. They start with a distribution of $16$ to $32$ embryos between $0.5$ and $1.5 \; \mathrm{AU}$ and do no include giant planets. Qualitatively, this is close to the state of our \ejs{} simulations after $10 \; \mathrm{Myr}$ when the gas has dissipated and inward sweeping secular resonances have set the final architecture of the system. We do not reproduce their observation that the number of planets decreases with the solid disk mass, likely because the number of embryos is set by sweeping secular resonances for us, whereas \cite{2006ApJ...642.1131K} treats this number as a free parameter. The simulations in \cite{2007ApJ...669..606R} use between $75$ and $190$ embryos, but also spread them out over a wider range of semi-major axes. They are therefore slightly more comparable to ours, but again do not include the effect of giant planets or assembly of embryos from planetesimals in the presence of gas.


\subsection{How Sweeping Resonances Sculpt the Disk}

In Sections \ref{sec:sResults_ssEvolution}, \ref{sec:sResults_ssArch}, and \ref{ssec:orbits}, we explored the key differences between our simulation runs. We found that the the presence, absence, or orbital configuration of the giant planets has the strongest influence on the forming terrestrial planets. In runs with giant planets, terrestrial planets form on tighter orbits, are less numerous, more massive, and dynamically hotter than in those without. In \ejs{} configurations, orbits are tighter and planets more massive and with slightly higher eccentricities and inclinations than in \cjs{} configurations. Changing the initial planetesimal distribution also had noticeable (but less drastic) effects. In particular, more massive initial planetesimals disks lead to more massive terrestrial planets. Steeper disks are more fickle, producing more massive planets on tighter orbits for \ejs{} runs, but the opposite in \cjs{} configurations. We now disseminate how the interactions between planetesimals, giant planets, and the massive gas disk drive these trends.

\subsubsection{Trapping Planetesimals in Resonances}

%
%

The bulk of the architectural properties of the final planetary systems are determined during the first few Myr when planetesimals -- embedded in a gaseous disk -- first grow into embryos through mutual collisions. At this point, their evolution is driven by their mutual gravitational interaction, their interaction with the gaseous disk, and with their interaction with the giant planets. Once the gas dissipates and the embryos have cleared out their immediate spheres of influence (a few Hill radii), few changes occur in the system architectures.

The most defining difference between simulations is the presence and orbital configuration of giant planets. Through mean motion and secular resonances, they can transfer significant angular momentum onto the planetesimals efficiently. This exchange also proceeds without resonances, but is less effective. Unless dampened, thusly excited planetesimals can be launched into the inner solar system or be ejected from the system. Planetesimals can also be repelled by or trapped in resonances. Those repelled are effectively blocked off from accessing regions of phase space. Planetesimals which are locked in resonances find their eccentricities and inclinations perpetually excited unless dampened.\footnote{In fact, the balance between dynamical excitement from a passing secular resonance and damping by hydrodynamic and gravitational interactions determines whether planetesimals are swept along in the first place \citep{2005ApJ...635..578N}.} Although these processes occur within the first few million years, they induce a signature of a dynamically hotter state which persists until the end of our simulations. Here, we find embryos and planetesimals in \ejs{} and \cjs{} runs at higher eccentricities and inclinations than those in \njs{} runs.

As illustrated in Section \ref{sssec:sweeping_resonances}, planetesimals that become trapped in secular resonances are swept along with the resonances, eventually settling near their final locations once the gas has dissipated ($\sim 10 \; \mathrm{Myr}$). In \ejs{} and \cjs{} configurations, the inward sweeping $\nu_5$ resonance delivers large amounts of material to its final location ($a \sim 0.9 \; \mathrm{AU}$). This significantly enhances the amount of material available for embryos to accrete from, thereby peaking the radial mass distribution. Planetesimals that manage to avoided being swept along by the $\nu_5$ resonance face the $\nu_6$ resonance soon after, which sweeps most of the remaining planetesimals inward, although some manage to avoid this fate. Although it does not drag planetesimals along, the $\nu_{16}$ resonance also sweeps outward \citep{2008ApJ...676..728T} and settles near $\nu_6$. Once the gas is gone, their location demarcates the outer boundary of the region populated by terrestrial planets. Those trapped in their vicinity find themselves excited onto eccentric orbits that either deliver them to the inner system to collide with embryos, fall onto the star, or be ejected from the system. After $147.84 \; \mathrm{Myr}$, most of the planetesimals beyond the $\nu_6$ and $\nu_{15}$ resonances have been removed, although stragglers remain. By being separated from the inner regions, they have avoided being accreted onto larger embryos, thus covering a wide range of masses. In systems without giant planets, on the other hand, no secular resonances exist, and planetesimals remain in region they are initially placed in (modulo inward drift due to hydrodynamic drag and gravitational scattering).

The net effect of inward sweeping $\nu_5$ and $\nu_6$ resonances is then to deliver planetesimals from the disk and trap them in the orbital region between. The region between these resonances is narrower in \ejs{} configurations that in \cjs{} ones. Naturally, this leads to a higher mass concentration in \ejs{} runs as well as a more complete conversion of planetesimals into planets, i.e. only a very weak bimodal signature in the mass-frequency distribution. Configurations with giant planets on circular orbits trap less mass per AU in the inner regions, which retards conversion of planetesimals into planets. As such, they retain a bimodal mass-frequency distribution after $147.84 \; \mathrm{Myr}$. As the conversion of planetesimals into embryos is more complete and sourced from a tighter orbital region with dynamically more excited planetesimals (two resonances are close-by), it is unsurprising to find more massive terrestrial planets in \ejs{} configurations. In systems with no giant planets, material is very much spread out, so that embryos only accrete slowly. This leads to low-mass terrestrial planets which are much more spread out in semi-major axis.

%
%

In disks with initially steep density profiles, sweeping resonances appear to be more efficient in shepherding material inwards than in the reference case. In this configuration, the systems are initialised with fewer planetesimals at larger semi-major axis so that (i) the resonances have to sweep less material along and (ii) the radial mass distributions are already biased towards smaller semi-major axes. The net result then is for systems with steep initial density profiles to have radial mass distributions and system architectures to appear almost indistinguishable from the reference disks. On the other hand, the initial bias of material to the inner regions also means that fewer planetesimals are dynamically excited by the sweeping resonances, resulting in a dynamically colder state with smaller eccentricities and inclinations.

If the initial planetesimal disk is massive, the dynamical evolution of planetesimals is more complicated. Although the initial excitement from resonances is similar to configurations with the reference and steep profiles, dampening by hydrodynamic drag is much less effective due to the larger planetesimal mass; see Figure 1 in \cite{2010Icar..207..517M}. This has three consequences. Firstly, sweeping secular resonances move fewer planetesimals inward.\footnote{If planetesimals have too large eccentricities, the damping timescale from interaction with the gas disk is longer than the time it takes for the secular resonance to sweep past the planetesimals. They cannot be trapped in the resonance \citep{2005ApJ...635..578N}.} Secondly, the richly excited planetesimals actually remove angular momentum from Jupiter through dynamical friction (e.g., \citet{2012PTEP.2012aA308K}), although this also happens to a lesser degree for disks following the reference and steep profiles. For example, the mean eccentricity of Jupiter during the first $312 \; \mathrm{kyr}$ is $e_\mathrm{J,Ref} = 0.0678$ and $e_\mathrm{J,Heavy} = 0.0738$ for \textsc{Run01} in the \ejs{} and \ejsheavy{} sets. After $10 \; \mathrm{Myr}$, the $312 \; \mathrm{kyr}$ averaged eccentricities are $e_\mathrm{J,Ref} = 0.0474$ and $e_\mathrm{J,Heavy} = 0.0385$, corresponding to a decrease of $43$ and $91$ per cent, respectively. Thirdly, a small population of planetesimals actually manages to move outwards past the orbits of Jupiter and Saturn. Although most are promptly ejected (on timescales of $\sim 10^5$ years), a single planetesimal (across twelve runs) is caught by Jupiter as a moon!\footnote{For videos and time-sliced figures of the planetesimals dynamics of all simulations, we refer the reader to the supplementary website.}

The final radial mass distributions and mass-frequency distributions in initially massive disks still resemble those of the reference configurations, which may be counter-intuitive. As sweeping resonances move fewer planetesimals, we would expect proportionally less mass to end up in the inner system. However, this appears to be amply compensated for by the larger mass of the planetesimals. The only striking remaining difference is that a larger amount of mass remains in the disk simply by virtue of starting out with more mass. Dynamically, the effect of more massive initial planetesimals is also somewhat opaque. While more massive planetesimals are more difficult to excite by sweeping resonances, their excitations are also more difficult to dampen by through hydrodynamic drag due to their size. Taking these considerations together, it appears that the more excited planetesimals are ejected from the system (or fall into the star) while the less excited ones are swept along into the inner system to build up terrestrial planets.

\subsubsection{Statistical Spread of Runs}

The dynamical outcome of a planetesimal passing through a resonance depends sensitively on the initial orbit. For two initially almost identical orbits, the more resonances the particles pass through, the farther their orbits can diverge. Therefore, simulations where particles are exposed to a larger number of resonances have more evolutionary pathways available. The wider spread of \ejs{} and \cjs{} runs with respect to \njs{} is then hardly surprising. By the same argument, we naively expect \ejs{} runs to have a larger spread than \cjs{} runs because more particles cross the sweeping $\nu_6$ resonance. While this holds at the 25/75 percentile level, it does not always hold at the 10/90 percentile level. Closer inspection of Figure \ref{fig:fig02_ae_snaps_res} reveals that in \cjs{} runs, the 3:1 and 5:2 mean motion resonances with Jupiter remains within the planetesimal disk, and the final location of the $\nu_6$ and $\nu_{16}$ (not shown) resonances are within $0.07 \; \mathrm{AU}$ of the 7:3 resonance. Over Myr timescales, such close stacking provides pathways for particle orbits to diverge and promotes ejection of bodies from the disk. As the effects of secular resonances and mean motion resonance cannot be readily distinguished, we caution from attributing to them any differences across \cjs{} and \ejs{} runs.


\subsection{The Spread of Typical Diagnostics (AMD, RMC)}
\label{ssec:spread_amd_rmc}

Simulations of terrestrial planet formation typically invoke diagnostics such as the Angular Momentum Deficit (AMD) and Radial Mass Concentration (RMC) to quantify how close they are to reproducing some observed reference system (which is usually the solar system). Due to their stochastic nature, we expect initially identical simulations to produce different values for the AMD and RMC. If values obtained from a given initial condition exhibit a large spread, they may overlap with values generated from a different initial condition. This makes it unclear whether the outcome arises from a different initial conditions or is merely a reflection of the stochastic nature of these simulations. This suggests that individual runs lack predictive power.

We now characterise the spread in AMD and RMC obtained in our simulations. The RMC is given by

\begin{equation}
  \mathrm{RMC} = \max \left( \frac{\sum_j m_j}{\sum_j m_j \left[ \log_{10} \left( a / a_j \right) \right]^2} \right),
\end{equation}

where the sum is over the mass $m_j$ and semi-major axis $a_j$ of the terrestrial planets, and we search for the maximum in a semi-major axis range covering all terrestrial planets \citep{1997A&A...317L..75L}. The AMD is

\begin{equation}
  \mathrm{AMD} = \frac{\sum_j m_j \sqrt{a_j} \left( 1 - \cos(i_j) \, \sqrt{1 - e_j^2} \right)}{\sum_j m_j \sqrt{a_j}},
\end{equation}

where $m_j$, $a_j$, $e_j$, and $i_j$ are the mass, semi-major axis, eccentricity, and inclinations of the terrestrial planets \citep{2001Icar..152..205C}.

\begin{figure}
\centering
\includegraphics[scale=0.82]{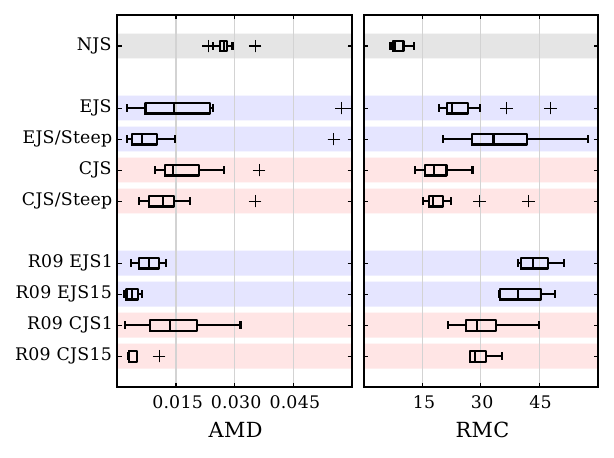}
\caption{Box plots of the Angular Momentum Deficit (AMD) and Radial Mass Concentration (RMC) obtained from (i) seven (out of nine) of our sets (\njs, \ejs, \ejssteep, \cjs, \cjssteep) and (ii) corresponding simulations from \citet{2009Icar..203..644R} (\textsc{R09 EJS1}, \textsc{R09 EJS15}, \textsc{R09 CJS1}, \textsc{R09 CJS15}). For visual aid, we indicate similar giant planet configurations through the background colour. Boxes indicate the $25/75$ range with the median marked in the box. Whiskers extend to $1.5$ times the inter-quartile range and crosses indicate outliers.}
\label{fig:fig10_amd_rmc}
\end{figure}

Figure \ref{fig:fig10_amd_rmc} shows the AMD and RMC from seven of our nine simulation sets as well as corresponding simulations from \cite{2009Icar..203..644R} who have used similar initial surface density profiles as well as giant planet orbits. Box plots are generated from (i) twelve runs per set (our runs), or (ii) four runs per set (R09).

Deferring a comparison to the simulations of \cite{2009Icar..203..644R} to Section \ref{ssec:ssd}, we find that our simulations without giant planets have small spreads in the AMD and RMC values. Simulations including giant planets have larger spreads, especially in the AMD. Values derived from \ejs{} and \cjs{} sets also overlap, and only the overall statistics reveal a trend towards smaller RMC values in \cjs{} runs. Spreads in runs with a steeper planetesimal distribution tend to be smaller except for RMC in \ejssteep{} runs. Overall, there is significant spread in all diagnostics, which emphasises the need for a statistical approach. Running only one simulation per set, pathological cases with reversed trends could arise, leading us to draw potentially wrongful conclusions.


\section{Discussion}
\label{sec:discussion}

Based on the above results, we now address a serious caveat associated with the paradigm of sweeping secular resonances, demonstrate large statistical spreads in typical diagnostics used in terrestrial planet formation, and attempt a preliminary link of our numerical results to observations.


\subsection{Sweeping Resonances: Summary \& Caveats}

Above, we have reinforced the notion of sweeping secular resonances to drive the dynamics of planetesimal disks. Depending on the configuration of giant planets, the $\nu_5$, $\nu_6$, $\nu_{15}$, and $\nu_{16}$ secular resonances with Jupiter and Saturn sweep inward as the gas dissipates until they settle in at their final positions (the inner $\nu_{16}$ resonances ``falls into the star'', as it were). Based on their dynamics, the secular resonances end up truncating the planetesimals disk which constrains growth of embryos to particular regions.


These dynamics are in line with previous work on secular resonances (as outlined in the introduction), although we demonstrate that they extend to the dynamics of planetesimals instead of only those of embryos and asteroids (which are, respectively, more massive or massless as far as giant planets are concerned).



However, a caveat exists in all numerical implementations above (including ours) -- the feedback of the gaseous disk onto the giant planets is neglected \citep{2014IAUS..310..194R}. As such, the eccentricity and inclinations of the giant planets will be dampened even beyond the current transfer of angular momentum onto the planetesimals and embryos. This modifies the time evolution and strength of the sweeping resonances which in turn affects the final configuration of terrestrial planets. For our simulations, this means that we may expect the \ejs{} runs to become a little more like \cjs{} runs. In other words, we most likely overestimate the mass and underestimate the semi-major axes of the final planets. Future work should clearly take into account a more sophisticated prescription of the gaseous protoplanetary disk that feeds back onto both the giant planets as well as the planetesimals and embryos.







\subsection{Spread of RMC \& AMD in Solar System Formation}
\label{ssec:ssd}

In Section \ref{ssec:spread_amd_rmc}, we presented the spreads of the Radial Mass Concentration (RMC) and Angular Momentum Deficit (AMD) in our simulation runs. As alluded therein, we now extend this to compare our spreads to those reported in previous work.

So far, we have deliberately omitted reference to the solar system, but note that the AMD and RMC diagnostics are frequently used to assess how close a given model manages to reproduce the solar system. As such, any comparison to previous work that reports on the AMD and RMC inevitably compares our simulations to those aimed at reproducing the solar system. This warrants a short overview of available models.

Although our initial conditions are based on simulations for exploring the formation of the solar system, our goal was not to explicitly to assess their viability for this. They are in fact not very well suited. For example, three key constraints in the solar system are the comparatively low mass of Mars, the compactness of terrestrial planet spacing, and the small eccentricities and inclinations of the terrestrial planets. While CJS configurations generate low-eccentricity planets, they also spread out the terrestrial planets too much and produce too massive Mars-analogues. In EJS configurations, the terrestrial planets tend to be more concentrated and low-mass Mars-analogues are abundant, but they tend to be too eccentric. In terms of the RMC and AMD, CJS configurations roughly match constrains for the AMD but not for the RMC; and vice versa for EJS configurations.

Barring unsuccessful attempts to sculpt extended planetesimals disks with extremely eccentric giant planets \citep{2009Icar..203..644R} or excitement of terrestrial planets by a late migration of giant planets \citep{2009A&A...507.1053B,2012ApJ...745..143A,2013MNRAS.433.3417B}, the most suitable initial conditions for generating terrestrial planets (up to a factors of a few) fulfill AMD and RMC constraints of the solar system require an initial planetesimal disk of extending only between $0.7$ and $1.0 \; \mathrm{AU}$ \citep{2008ApJ...685.1247M,2009ApJ...703.1131H} instead of between $0.5$ and $4.0 \; \mathrm{AU}$. It remains unclear whether the outer edge of the annulus results from a truncation by migration of giant planets \citep{2011Natur.475..206W} or is simply indicative of the initial mass distribution of solids in the protosolar nebula \citep{2002M&PS...37.1523C,2008ApJ...674L.105J,2014ApJ...782...31I}. The inner edge may be sculpted by material being trapped and piling up in a pressure bump \citep{2004ApJ...601.1109Y,2009ApJ...697.1269J}. For a more thorough exploration of initial conditions suitable to reproduce the solar system terrestrial planets, we refer to \cite{2014ApJ...782...31I}, \cite{2014ApJ...794...11I}, \cite{2011Natur.475..206W}, \cite{2010Icar..207..517M}, \cite{2009Icar..203..644R}, \cite{2006ApJ...642.1131K}, and \cite{2001Icar..152..205C} as well as the review of \cite{2014prpl.conf..595R}.

A number of authors report AMD and RMC for their solar system formation simulations, but have only run a single instance for each set of simulation parameters \citep{2001Icar..152..205C,2010Icar..207..517M}. To our knowledge, only \cite{2006ApJ...642.1131K}, \cite{2009Icar..203..644R}, as well as \cite{2014ApJ...782...31I} account for stochastic variations across runs by running similar initial conditions multiple times.\footnote{In contrast to our simulations, \cite{2006ApJ...642.1131K} and \cite{2009Icar..203..644R} do not evolve identical initial conditions, but rather redraw different initial conditions from the same underlying distribution. We are uncertain what is done in \cite{2014ApJ...782...31I}.}$^,$\footnote{\cite{2003AJ....125.2692L} and \cite{2004Icar..168....1R} also report stochastic variations across simulations, but do not report the RMC or AMD due to their focus on different aspect.} Of these, \cite{2009Icar..203..644R} (hereafter R09) consider and present simulations most similar to our own, and we include some of their results here for comparison; cf. Figure \ref{fig:fig10_amd_rmc}.

Although the ranges in AMD and RMC populated by our runs and those of R09 tend to be misaligned, they generally show overlap. More importantly, at least some trends resulting from changing the orbits of the giant planets and initial planetesimal mass distribution appear to robust. For example, in CJS runs, moving to steeper initial planetesimal distribution decreases the AMD while the RMC remain approximately the same. For EJS runs, steeper initial profiles also decrease the AMD in both cases, although the RMC responds differently in the runs of R09 vis-\`a-vis our runs. Changing the orbits of the giant planets is much more robust and exhibits the same trends in both sets (when judging by the median).

Overall, it appears that runs in R09 follow clear trends and are not as burdened by overly wide spreads and outliers as ours. Unfortunately, this clarity appears to be a consequences of simplifications made to conform to the computational resources available at the time. In particular, recall that depending on the giant planet configuration and initial planetesimal distribution, inward sweeping secular resonances generate a particular distribution of embryos after the gas disk has dissipated. It is only at this point that R09 initialise their simulations, by necessity with a more generic embryo distribution. This distribution lacks the imprint of the important dynamics that took place as the gas dissipated. As much of the shaping of the final system has already happened at this point, it should not be surprising that their runs produce smaller spreads in diagnostics.\footnote{Of course, we have been wholly ignorant about the fact that giant planets may not yet be present (or massive enough) on their orbits as we start our runs. Future work will likely improve upon this and cheerfully point out this shortcoming.}

It appears that modelling of the initial phase of planetesimals embedded in a dissipating gas disk is essential. Omitting this stage neglects much of the dynamics that shape the final architecture of the system.


%
%

\subsection{Preliminary Comparison to Observations}

\begin{figure}
\centering
\includegraphics[scale=0.86]{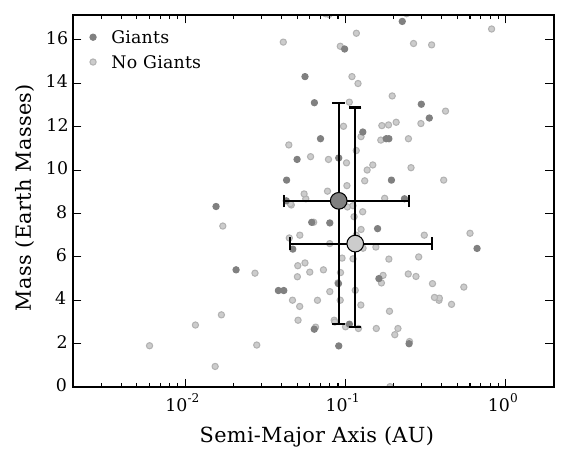}
\caption{Mass and semi-major axis for all sub-Neptunes ($M < 17.14 \; \Mearth$) in the reduced (see text) sample of observed exoplanets. Small points are observational data. Light and dark grey points are planets in systems with and without detected giant ($M \geq 50 \; \Mearth$) planets in the same system. Large circles and lines indicate the median as well as $10/90$ percentiles for the two sets.}
\label{fig:fig11_observations}
\end{figure}

At this point, we may wonder whether the key trends found in our simulations also hold in extrasolar planetary systems. In other words, are the observed terrestrial planets\footnote{We adopt working definitions of terrestrial and giant planets as planets with masses $M < \Mneptune \sim 17.15 \; \Mearth$ and $M > 50 \; \Mearth$.} more massive, less numerous, and closer to the host star if the system also hosts giant planets? Deferring caveats of our admittedly naive comparison for now, Figure \ref{fig:fig11_observations} shows the mass and semi-major axis of all sub-Neptune mass planets found in multiple planet systems, grouped into whether the system also hosts giant planets or not. We also summarise their statistical properties in Table \ref{table:observations}. Exoplanet observations are loaded from the \textit{Extrasolar Planet Encyclopaedia}\footnote{\url{http://www.exoplanet.eu}} database \citep{2011A&A...532A..79S} using a snapshot from 02 August 2015. We only consider systems hosting at least two terrestrial planets and require complete data on semi-major axis and mass for all planets in system. Raw data from a total of $1228$ planetary systems is reduced to $62$ systems, $43$ of which host giant planets.

\renewcommand{\arraystretch}{1.3}
\begin{table}
\caption{Median and $10/90$ percentiles of mass $M$, semi-major axis $a$, and number of planets $N_P$ for the sub-Neptune ($M < \Mearth$) population of two sets of exoplanet systems. The number of system fulfilling the filtering criteria is $N_S$, and $f_S$ the corresponding fraction out of $1228$ total systems.}
\label{table:observations}
\begin{center}
\begin{tabular}{c c c c c c}
\toprule

Set & $M$ ($\Mearth$) & $a$ ($\mathrm{AU}$) & $N_P$ & $N_S$ & $f_S$ (Per cent)\\
\midrule
No Giants & $6.61_{-3.84}^{+6.27}$ & $0.12_{-0.07}^{+0.23}$ & $2_{-1}^{+1}$ &$43$ & $3.5$ \\
Giants & $8.58_{-5.68}^{+4.52}$ & $0.09_{-0.05}^{+0.16}$ & $1_{-0}^{+2}$ &$19$ & $1.5$ \\

\bottomrule
\end{tabular}
\end{center}
\end{table}
\renewcommand{\arraystretch}{1.0}

Overall, the observations appear to corroborate the results of our simulations. In systems hosting giant planets, the median terrestrial planet mass is $30$ per cent larger, the median semi-major axis $25$ per cent smaller, and the terrestrial planets are generally less numerous. Although encouraging, these results clearly suffer from two major caveats. First, the observed planets are much more massive than those in our simulations. Second, the observed distribution of planets peaks around $0.1 \; \mathrm{AU}$ while our simulations concentrate planets at about $1.0 \; \mathrm{AU}$.

Superficially, the differences in planetary masses can be resolved by imposing that the observed planets formed in a more massive initial solid disk. In Section \ref{sec:results}, we found that doubling the initial mass in planetesimals tends to increase the median mass of the final terrestrial planets by about $50$ per cent. By this measure, we would require the initial protoplanetary disks to host about $80 \; \Mearth$ of solid material (four doublings), although much larger required enhancements have also been suggested \citep{2014ApJ...795L..15S}. This does not account for the accretion of gas onto a rapidly formed core, which would account for some of the additional mass. A steeper initial surface density profile can further decrease the required total mass in solids throughout the disk by concentrating more material in the inner regions. However, surface density profiles steeper than $\Sigma_\mathrm{Solids} \varpropto r^{-1.5}$ are not supported by observations \citep{2008ApJ...678.1119H,2009ApJ...701..260I,2009ApJ...700.1502A,2010ApJ...723.1241A,2011ARA&A..49...67W}.

A more massive initial disk and steeper surface density profile would significantly shorten the formation timescale of terrestrial planets. This especially holds for those that may have formed in-situ, i.e. at their detected orbital location around $0.1 \; \mathrm{AU}$. Absent gas and giant planets, the timescale for a planet to grow to isolation mass scales roughly as $t_\mathrm{Isolation} \varpropto a^3 / (\Sigma_\mathrm{Solids})^{1/2}$ \citep{1990Icar...88..129I,2012PTEP.2012aA308K}. In other words, while increasing the total solid disk mass by four doublings shortens the formation timescale by only a factor of four, in-situ formation at $0.1 \; \mathrm{AU}$ (instead of at $1.0 \; \mathrm{AU}$) accelerates accretion of planetary embryos by a factor $1000$. In this case, terrestrial planets would have formed much faster than the giant planets, which implies that their inward sweeping secular resonances would have little effect on the already formed planets by virtue of their large mass.

On the other hand, planetesimals  accrete while being immersed in the gaseous disk. Depending on the disk structure, the presence or absence of giant planets, the gas dissipation rate, and the mass of the forming embryos, radial migration of planetesimals, embryos, and formed terrestrial planets is an almost inescapable consequence \citep{2005ApJ...635..578N,2006ApJ...652..730M,2007ApJ...654.1110T,2008ApJ...676..728T,2008A&A...482..677C,2010MNRAS.401.1691M,2014A&A...569A..56C}. Planetesimals move inwards either from hydrodynamic drag, secular resonance sweeping, or shepherding while migration of embryos and planets is mostly driven by disk interactions. This scenario is conceptually much more in line with our simulations and provides sufficiently long formation timescales for giant planets to affect the inner planets as they form.


In either scenario, embryos and terrestrial planets migrate inward as they interact with the dissipating gas disk with the rate and direction of migration sensitively depending on the structure of the gas disk \citep{2010MNRAS.401.1950P,2011MNRAS.410..293P,2013A&A...549A.124B,2014A&A...564A.135B,2014A&A...570A..75B}. Typically, migration terminates at the inner edge of the disk ($\sim 0.1 \; \mathrm{AU}$) where planets tend to pile up in resonant chains \citep{2007ApJ...654.1110T,2008MNRAS.384..663R,2009ApJ...699..824O,2014A&A...569A..56C}, although this is inconsistent with observed period ratios \citep{2011ApJS..197....8L}. In our simulations, we remove planets approaching to within $0.2 \; \mathrm{AU}$ of the host star to meet integration accuracy requirements. This means that we potentially neglect a population of planets interior $0.2 \; \mathrm{AU}$. To test for this, we have carried out additional test runs (at lower accuracy) where we only remove material that approaches the host star to within two Solar radii ($\sim 0.005 \; \mathrm{AU}$). In these tests, we indeed observe a clustering of planets around the inner edge of the gas disk, which we have placed at 0.1 AU.\footnote{We consider the gaseous disk to extend inwards to $0.1 \; \mathrm{AU}$, which corresponds to the dust sublimation radius \citep{2007prpl.conf..555D}. Of course, that does not necessarily mean that the gas disk extends to the same inner radius. To within a factor of two, observations indicate the inner edge of the gaseous disk to coincide with the Solar corotation radius \citep{2007prpl.conf..507N}, which is at $0.17 \; \mathrm{AU}$ for the present-day Sun.} As such, we are conceptually able to reproduce the observed mass distribution of extrasolar terrestrial planets.

In these trials, however, the formation of planets interior to $0.2 \; \mathrm{AU}$ appears to only be loosely coupled to the evolution of the giant planets. Initially, they grow solely by accretion of local material while migrating inwards until their eventual pile-up at inner edge of the gaseous disk. Compared to embryos beyond $0.2 \; \mathrm{AU}$, they grow faster, have a smaller final mass,\footnote{Within the first $\mathrm{Myr}$, the $90$ percentile mass of embryos located within $0.2 \; \mathrm{AU}$ rapidly grows to $\sim 0.25 \; \Mearth$, where growth stalls out. For embryos beyond $0.2 \; \mathrm{AU}$, the $90$ percentile mass reaches this threshold between $4$ to $5 \; \mathrm{Myr}$, but these embryos keep accreting afterwards.} and source comparatively small amounts of material from the regions affected by inward sweeping resonances.\footnote{By $\sim 7.7 \; \mathrm{Myr}$, $95$ per cent of material interior to $0.2 \; \mathrm{AU}$ originates from within $1.35 \; \mathrm{AU}$, whereas embryos in the region $0.2 < a < 1.0 \; \mathrm{AU}$ source over $95$ per cent of their material from beyond $1 \; \mathrm{AU}$.}

Given the the differences in mass and distribution of terrestrial planets between our simulations and observations, a direct comparison between the two may not be well-justified as yet. Nevertheless, it is encouraging that basic trends persist, such that we expect more suitable simulations to be able to match the basic properties of at least some of the exoplanetary systems discussed above. In such a scenario, we also expect to be able to predict as yet undetected giant planets (and their orbital configuration) by considering outliers in the system architectures of the terrestrial (sub-Neptune mass) planets. For example, if the terrestrial planets in a system without detected giant planets are significantly more massive, less numerous, and closer-in than the median values for this population, we may expect undetected giant planets to be present. A cursory exploration of this scenario suggests that CoRoT-7, HD 20003, and HD 20781 may host undetected giants. These systems should be the first targets in a future analysis.


\section{Summary \& Conclusions}
\label{sec:conclusions}

In this work, we have addressed the stochastic nature of terrestrial planet formation. Orbits which are initially identical up to one part in $10^{15}$ ($\sim$ double precision floating point accuracy) diverge exponentially until their separation becomes limited by the accessible phase-space. The divergence is driven by differences in the geometry of successive close encounters with other particles. We find that the rate of divergence increases with the number of simulation particles. After a few hundred years, orbits have diverged far enough for the collisional history of two simulations with initially nearby orbits to differ. They are now fully diverged, and produce distinctly different planetary systems within our simulation times of $\sim 147.84 \; \mathrm{Myr}$.

If the position of a single planetesimal is changed by less than one millimetre, the positions, orbits and masses of the resulting planets are all different. We find that nearby orbits diverge exponentially with an e-folding timescale of a few to a few tens of years which decreasing as the number of particles increases. There is no reason to expect that this behaviour does not continue to much smaller scales. Perhaps if our early solar system had contained one extra molecule, the Earth would not have formed at all.

Even if simulations are initialised with identical initial conditions, variations in round-off errors quickly seed differences in the planetesimal orbits, which are amplified by sequences of close encounters. The variations are caused by the ill-defined behaviour of certain intrinsic functions available in parallel programming. They cause loss of control over the order of operations in the code. Although this can be mitigated, the extreme sensitivity on the order of operations implies the existence of multiple --  numerically equally valid -- final configurations for any given initial condition. Each of these configurations can give wildly different values for typical diagnostics. For example, we find variations in the final surface density profiles and note that the final profiles are not correlated with the initial ones. Individual simulations therefore lack predictive power, and we are relegated to considering distributions and stacks of multiple runs. Unfortunately at present, it remains unclear how many runs are required to adequately sample the systems. We advocate at least $8$ runs per initial condition. This would generate two samples per quartile given an underlying uniform distribution.

Analysing our simulations, we find that varying the configuration of giant planets in systems has statistically robust results in line with previous work. Planetesimal dynamics are driven primarily by sweeping secular resonances and the extend of the terrestrial planet is set by the final locations of the resonances. Systems with giant planets tend to form fewer, more massive, and more eccentric terrestrial planets at smaller semi-major axes than those without. Although we probe different mass and orbital regimes, observations of sub-Neptune mass planets appear to support these trends. A proof-of-concept extraction of outliers in the observations suggests that CoRoT-7, HD 20003, and HD 20781 may host as yet undetected giant planets.

\section*{Acknowledgements}

We are most grateful to the anonymous referee as well as Sean Raymond for a host of useful comments that improved the quality of this manuscript. We also wish to thank Ryuji Morishima, Rok Ro\v{s}kar, Thomas Peters, George Lake, Michael Rieder, and Joanna Dr\c{a}\.{z}kowska for useful discussions, as well as Doug Potter for computational support. Simulations were run on the Tasna and Vesta GPU clusters at the Institute for Computational Science, University of Z\"urich. We also acknowledge use of Numpy, Scipy, IPython/Jupyter, Matplotlib, and Colorbrewer2 for post-processing.

\bibliographystyle{mnras}
\bibliography{chaos}

\bsp

\label{lastpage}

\end{document}